\documentclass[journal,twoside,web]{ieeecolor}
\usepackage{generic}
\usepackage{cite}
\usepackage{flushend}
\usepackage{amsmath,amssymb,amsfonts}
\usepackage{algorithmic}
\usepackage{graphicx}
\usepackage{textcomp}
\usepackage{comment}
\usepackage{caption}
\usepackage{subcaption}
\usepackage{float}
\usepackage[export]{adjustbox}
\usepackage{tcolorbox}
\usepackage{hyperref}
\usepackage{tikz}
\usetikzlibrary{spy,calc}
\def\BibTeX{{\rm B\kern-.05em{\sc i\kern-.025em b}\kern-.08em
    T\kern-.1667em\lower.7ex\hbox{E}\kern-.125emX}}
\begin{document}
\title{Cross-modal tumor segmentation using generative blending augmentation and self-training}
\author{Guillaume Sallé, Gustavo Andrade-Miranda, Pierre-Henri Conze, Nicolas Boussion, Julien Bert, \\Dimitris Visvikis$^\dagger$, Vincent Jaouen$^\dagger$
\thanks{The authors are with LaTIM, UMR Inserm 1101, Université de Bretagne Occidentale, IMT Atlantique, Brest, France.}
\thanks{$^\dagger$ D. Visvikis and V. Jaouen share senior authorship.}
\thanks{N. Boussion and J. Bert are also with CHRU Brest University Hospital, Brest, France.}}
\maketitle

\begin{abstract}

\textit{Objectives}: Data scarcity and domain shifts lead to biased training sets that do not accurately represent deployment conditions. A related practical problem is cross-modal image segmentation, where the objective is to segment unlabelled images using previously labelled datasets from other imaging modalities. 
\textit{Methods}: We propose a cross-modal segmentation method based on conventional image synthesis boosted by a new data augmentation technique called Generative Blending Augmentation (GBA). GBA leverages a SinGAN model to learn representative generative features from a single training image to diversify realistically tumor appearances. 
This way, we compensate for image synthesis errors, subsequently improving the generalization power of a downstream segmentation model. The proposed augmentation is further combined to an iterative self-training procedure leveraging pseudo labels at each pass.
\textit{Results}:  The proposed solution ranked first for vestibular schwannoma (VS) segmentation during the validation and test phases of the MICCAI CrossMoDA 2022 challenge, with best mean Dice similarity and average symmetric surface distance measures.
\textit{Conclusion and significance}: Local contrast alteration of tumor appearances and iterative self-training with pseudo labels are likely to lead to performance improvements in a variety of segmentation contexts. 
\end{abstract}


\section{Introduction} 
\label{sec:introduction}

\IEEEPARstart{I}{n} machine learning (ML) for medical image analysis, a critical hypothesis is that training and testing images are drawn from the same data distribution. However, factors such as varying clinical centers, population, imaging modalities, acquisition protocols or vendors, make training sets unrepresentative and biased compared to deployment conditions~\cite{guan2021domain}. This domain shift is a major limitation to the use of learning-based models in medical imaging. A related issue is data scarcity: when training data are scarce, the sample selection bias further increases the domain shift~\cite{glocker}. Therefore, there is a clear need to develop models that can  be both robust to out-of-distribution samples and also generalize from low training data regimes. 

\begin{figure}[!t]
\begin{center}
\includegraphics[width=0.95\linewidth]{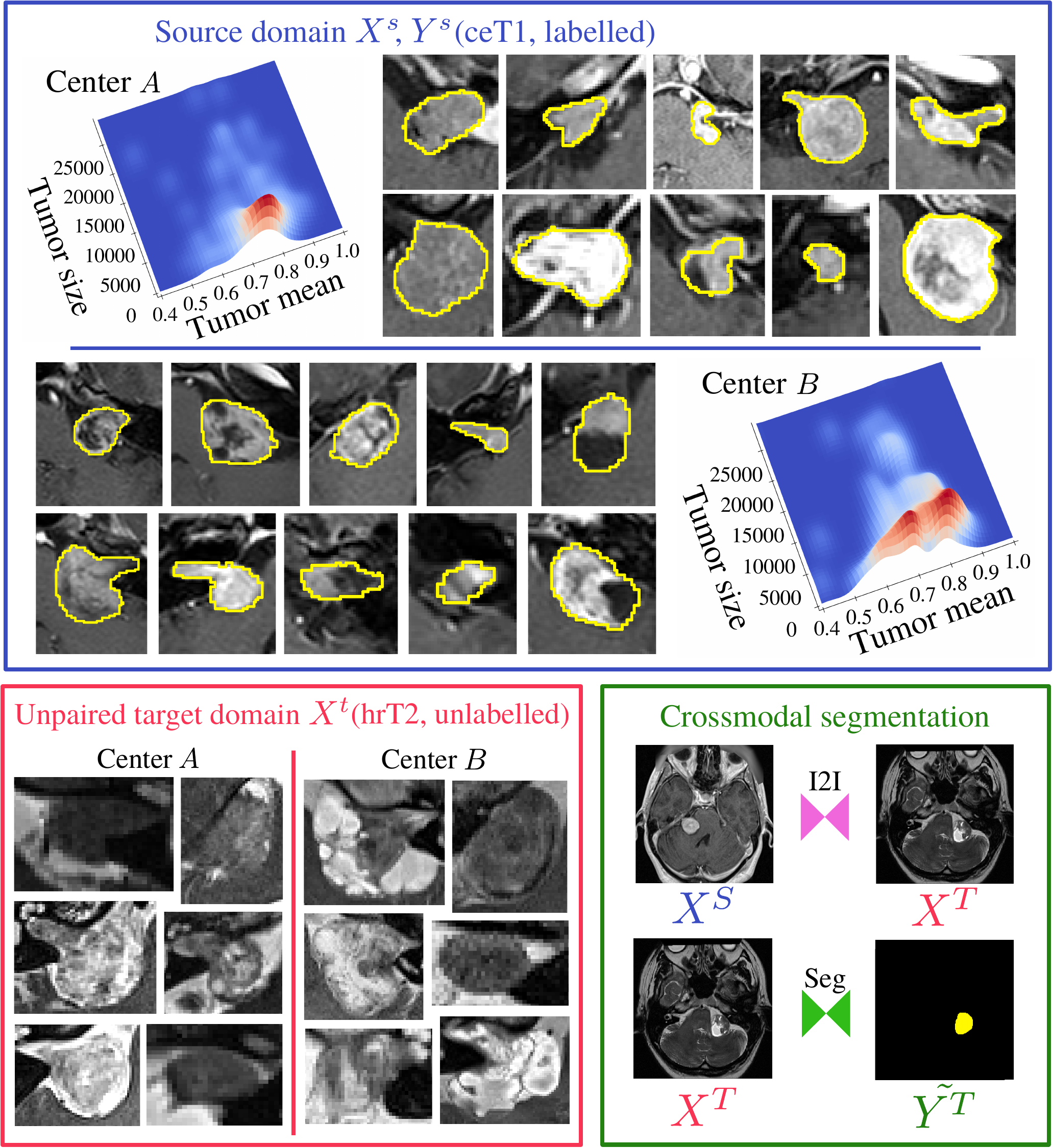}
\caption{Example cross-modal tumor segmentation scenario (CrossMoDA 2022 challenge). Source domain images $X^S$  (blue frame, here contrast-enhanced T1 MRI) are provided with segmentation labels $Y^S$, while target domain images $X^T$ (red frame, here high-resolution T2 MRI) are unlabelled. Tumors show variable visual appearance between two imaging centers, with variations in modes of intensity and size (shown in kernel density plots). 
} \label{fig1}
\end{center}
\end{figure}

A  clinical challenge related to these limitations is to perform \textit{unsupervised cross-modal segmentation}, i.e. to exploit  previously annotated source data on new, unlabelled target domains~\cite{dorent2020scribble}. For instance, cross-modal segmentation could prove useful in the context of vestibular schwannoma (VS) segmentation, a benign tumor of the vestibular nerve sheath, which is the central focus of the MICCAI CrossMoDA 2022 challenge~\cite{CrossMoDA2021}.  
In most current clinical workflows for radiation therapy treatment planning of VS, contrast-enhanced T1-weighted (ceT1) pulse sequences is used to identify and outline the tumor volume (i.e. the VS), while the delineation of the organ at risk cochlea is obtained from dedicated high-resolution T2 (hrT2) sequences~\cite{vokurka2002using,shapey2021segmentation}. In order to reduce the use of gadolinium contrasting agents and to simplify patient management, there has been significant interest in exploring the potential use of high-resolution T2 (hrT2) sequences for VS segmentation~\cite{coelho2018mri,wang2019automatic}. However, substantial domain gaps prevent the direct reuse of methods trained on ceT1 images on hrT2 images, which is further increased by changes in protocols and vendors across clinical centers. Fig.~\ref{fig1} illustrates the variability of tumor appearances in ceT1 images. The tumors either exhibit positive or negative contrast with respect to surrounding tissues and present a diversity of textural heterogeneity across subjects and centers. A typical cross-modal solution involving image-to-image translation followed or combined to segmentation (green frame) may fail to capture this variability. 

In this work, we present two key contributions aiming at improving the generalization power of downstream segmentation models in conventional cross-modality pipeline involving image-to-image (I2I) translation followed by segmentation: 
\begin{itemize}
  \item We introduce a new data augmentation technique to diversity tumor appearance to which the network is exposed during training. To this end, we analyze the distribution of tumor appearances in the target domain to cover the distribution shift between centers by leveraging pseudo labels achieved through iterative self-training. Doing so, we make the segmentation model more robust to center-specific features and potential errors from the I2I translation stage.
  
  \item We further blend the synthetically altered tumor using a SinGAN model~\cite{Shaham2019} based on a cascade of generators trained on a single image slice to improve segmentation performance using iterative self training, a technique we call Generative Blending Augmentation.

\end{itemize}

Our solution combined to an iterative self training scheme obtained the best average VS Dice score and was ranked first during both the validation and test phases of the CrossMoDA 2022 challenge for the VS segmentation subtask according to the ranking system proposed by the organizers. We achieved an overall third rank when combined to the cochlear segmentation task, not addressed by our method. {Our challenge docker image and technical details regarding execution are publicly available at \footnote{\url{https://github.com/GuigzoS/crossmodalGBA}}.}
 

\section{Related Works}

In this section, we discuss works related to our approach directed towards improving the generalization power of segmentation models through unsupervised domain adaptation and data augmentation.

\subsection{Cross-modal segmentation}

Deep domain adaptation (DA) methods are being increasingly studied in medical image segmentation to reduce the domain shift effects~\cite{guan2021domain,tajbakhsh2020embracing,wang2018interactive}.  In the context of cross-modal segmentation, we focus in particular on unsupervised domain adaptation (UDA) methods that do not rely on any prior knowledge of the labels of the target domain~\cite{perone2019unsupervised,zou2018unsupervised, hognon2019standardization,tixier2021evaluation,hognon2023contrastive}.  Typically, UDA methods for cross-modal segmentation involve two stages: unsupervised image-to-image (I2I) translation to learn intensity mappings between source and target domains, followed by supervised segmentation leveraging labels from the source domain~\cite{cosmos,fu2020domain,zhao2021mt,zeng2021semantic,hoffman2018cycada,zhao2023ms}. These two stages can also be combined into end-to-end models to benefit from label knowledge during I2I translation, at the cost of increased architectural complexity~\cite{cosmos,huo2018synseg,chen2020unsupervised}.  Although newer generative paradigms based on e.g. diffusion models are now emerging~\cite{ozbey2022unsupervised}, most of existing I2I translation methods are based on generative adversarial networks (GANs)~\cite{guan2021domain,CrossMoDA2021,zhao2022uda,cosmos} that promote realistic outputs through a competition between a generator and a discriminator~\cite{jiang2018tumor,fard2022cnns}. The most popular models for unsupervised I2I translation are CycleGAN and its variants~\cite{cyclegan, park2020contrastive, chen2020reusing}. However, GAN-based methods tend to learn global image-level mappings, potentially disregarding smaller regions of interest (ROIs) like tumors that may be underrepresented in the training set~\cite{cohen2018distribution, cohen}.  Maintaining a balance in the distributions of features of interest in the training and target domains becomes crucial for accurately translating such structures, which is of paramount importance to train a downstream segmentation model on the target modality. Tuning this proportion without prior knowledge of the test set's composition remains an open problem. Lastly, an often overlooked aspect is the high variability and difficulty in reproducing the outputs of CycleGAN models~\cite{huang2022stability, liu2020illumination}. It is common to retain the best performing model (as measured subjectively) from several trainings, which is not satisfactory due to the aleatoric nature of such practice.

\subsection{Data augmentation}

Data augmentation allows to artificially increase the diversity of training examples without additional data~\cite{chen2022generative}. 
Conventional methods include simple image manipulation such as scaling, cropping, rotation, flipping or global contrast change~\cite{shorten2019survey,ronneberger2015u}. More complex methods can also be considered, such as e.g. mix-up to generate convex combinations of training samples~\cite{zhang2017mixup,cao2022scoremix,panfilov2019improving}. While these techniques, originally designed for natural image processing, may stabilize training to some extent, they are not particularly designed to the properties of medical images. Consequently, there is a growing focus on exploring augmentations specifically tailored to the field of medical imaging~\cite{nalepa2019data,zhao2019data,chlap2021review,kebaili2023deep}. For instance, some approaches propose to generate realistic artefacts or plausible intensity variations through style transfer~\cite{chen2020realistic,hesse2020intensity}. These methods may improve the model's performance and robustness, but are not designed to reduce an observed domain shift with respect to a given dataset.

Due to their generative capability, GANs can also be trained on medical images to generate samples tailored to a specific modality~\cite{osuala2022data}. Related methodologies include tumor inpainting techniques from healthy subjects~\cite{kim2021synthesis,li2020tumorgan,micgs2022} or medical image style transfer~\cite{micgs}. Such GAN-based data augmentations have demonstrated a potential to reduce the number of required annotations compared to traditional data augmentation methods. However, these approaches typically rely on large training datasets to enable the generator and the discriminator to capture the diversity of the data distribution. Consequently, they lack robustness in scenarios with limited data availability.



\begin{figure}[ht]
\begin{center}
\includegraphics[width=\linewidth]{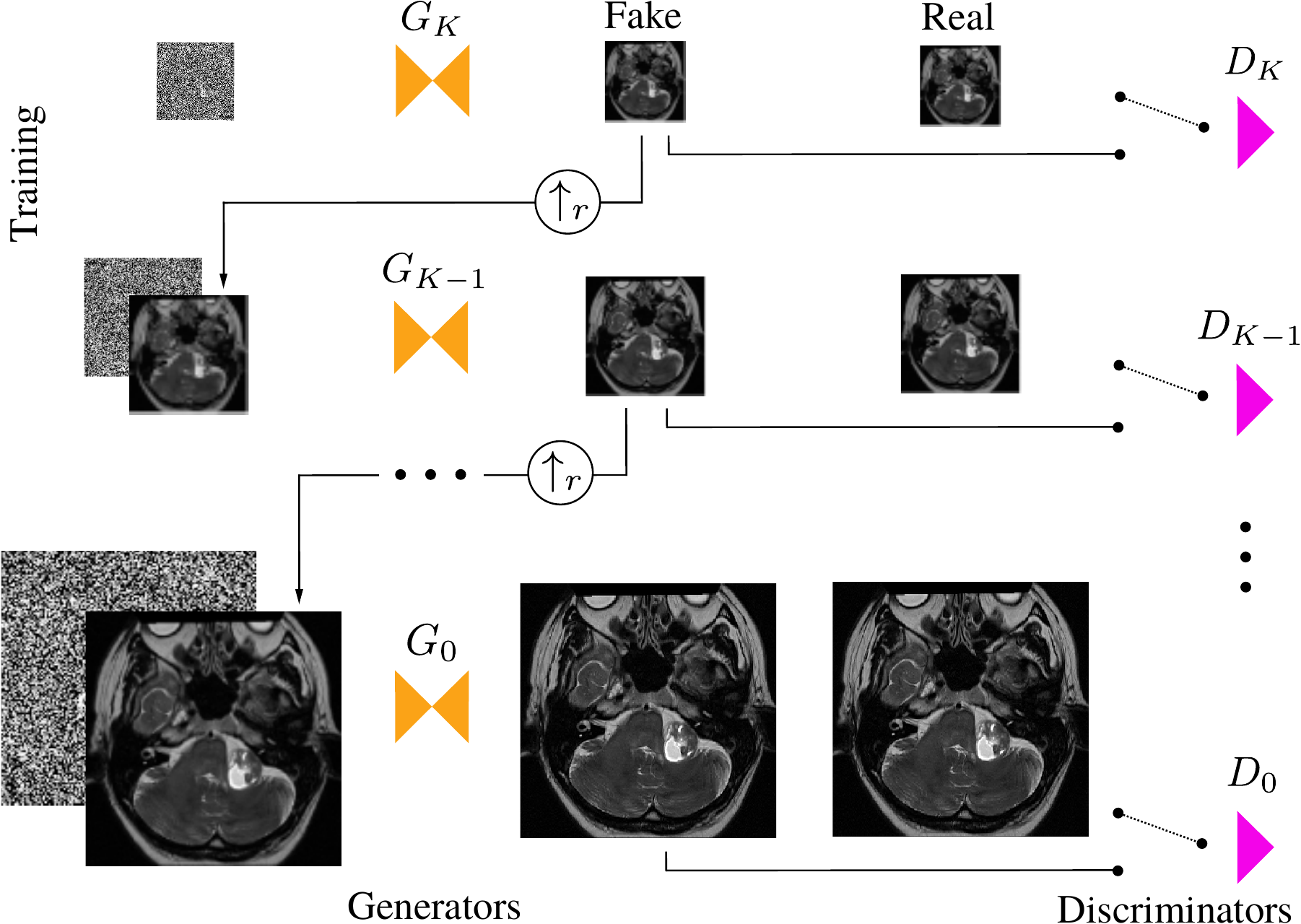}
\caption{SinGAN generative model from a single 2D image \cite{Shaham2019}. Each generator/discriminator pair $(G_k, D_k)$ is trained independently and successively, from coarsest ($k=K$) to finest scale ($k=0$). Scale level is obtained by downsampling the original image by a factor $r^k$.  }\label{singan}
\end{center}
\end{figure}

\section{Method}

In this section, we present our proposed approach for unsupervised cross-modal segmentation. It consists of a conventional cross-modal pipeline (i.e. I2I translation, then segmentation) where we focus on improving the generalization power of the segmentation model through a novel data augmentation technique and iterative self-training. 

\subsection{Notations and general objective}

Let $(X^S, Y^S)$ be the source ($S$) image set with associated tumor labels (for example ceT1-MRI)  and $X^T$ the target ($T$) image set (for example hrT2-MRI). Each modality set may be composed of images coming from multiple centers. In this work, without loss of generality, we restrict the problem to the case of two clinical centers $A$ and $B$, i.e. $X^S = X_A^S \cup X_B^S$ and $X^T = X_A^T \cup X_B^T$. Note that there may exist potentially large domain shifts between $A$ and $B$ due to the center effect, making it impossible to learn a meaningful $S\rightarrow T$ translation when both centers are used simultaneously. Our global objective is to generate a realistic set of tumor annotations $Y^T$ to train a segmentation model downstream on the target image set $X^T$.

\begin{figure*}[!t]
\centering
\includegraphics[width=\textwidth]{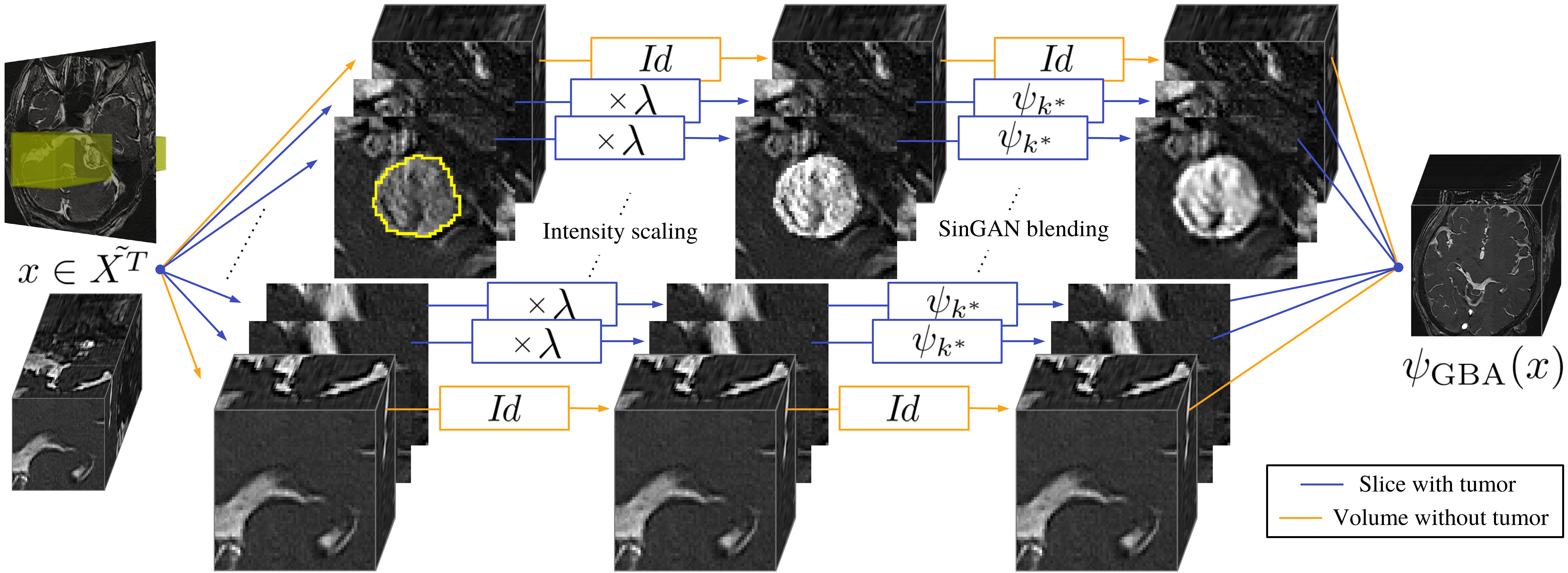}
\caption{Generative blending augmentation (GBA) of a vestibular schwannoma in a synthetic hrT2 image generated by CycleGAN (CrossMoDA 2022 dataset). For each {tumor slice}, tumor intensity values are linearly scaled by a factor $\lambda$ to perform contrast alteration. A SinGAN generative model is then used to realistically blend the contrast-altered tumor to the original image. The cascaded SinGAN generators are trained on the central slice and applied to every slice containing the tumor. }\label{workflowpart1}
\end{figure*}

\subsection{Image-to-image translation}

First, we seek to learn a pixel mapping from between the source modality $X_A^S$ and the target modality $X_A^T$ using a CycleGAN  unsupervised I2I translation model. We restrict the training to one of the two centers (here $A$), due to internal (i.e. intra-modality, cross-center) domain shifts that would prevent meaningful mappings between modalities if trained globally using data from centers $A$ and $B$. We train two generators $({\mathcal G}_{S \to T},{\mathcal G}_{T \to S})$ and two  discriminators $(\mathcal D_S, \mathcal D_T)$ under a combined adversarial and cycle-consistent objective. The adversarial loss is defined as: 




\begin{equation}
\begin{split}
\mathcal{L}_{adv}(& {\mathcal G}_{S \to T} , {\mathcal D}_T, X^S, X^T) =\mathbb{E}_{x^T \sim X^T}[log({\mathcal D}_T(x^T))] \\ 
&+ \mathbb{E}_{x^S \sim X^S}[log(1-{\mathcal D}_T({\mathcal G}_{S \to T}(x^S)))]
\label{eq:eqADV}
\end{split}
\end{equation}
and the cycle-consistent loss as:
\begin{equation}
\begin{split}
\mathcal{L}_{cyc}( {\mathcal G}_{S \to T} &, {\mathcal G}_{T \to S}, X^S, X^T) = \\
& \mathbb{E}_{x^S \sim X^S} [\| {\mathcal G}_{T \to S}({\mathcal G}_{S \to T}(x^S)) - x^S \|_1] \\
+ & \mathbb{E}_{x^T \sim X^T} [\| {\mathcal G}_{S \to T}({\mathcal G}_{T \to S}(x^T)) - x^T \|_1].
\label{eq:eqCYC}
\end{split}
\end{equation}
The full objective function is then:
\begin{equation}
\begin{split}
\mathcal{L}( {\mathcal G}_{S \to T} &, {\mathcal G}_{T \to S}, {\mathcal D}_S, {\mathcal D}_T, X^S, X^T) = \\
& \mathcal{L}_{adv}( {\mathcal G}_{S \to T} , {\mathcal D}_T, X^S, X^T) \\
+ & \mathcal{L}_{adv}( {\mathcal G}_{T \to S} , {\mathcal D}_S, X^T, X^S) \\
+ & \mu_{cyc} \cdot \mathcal{L}_{cyc}( {\mathcal G}_{S \to T}, {\mathcal G}_{T \to S}, X^S, X^T)
\label{eq:eqI2I}
\end{split}
\end{equation}
Only ${\mathcal G}_{S \to T}$ is then used to generate synthetic images $\tilde{X^T}$ for the target domain: 
\begin{equation}
\tilde{X^T}={\mathcal G}_{S \to T}(X^S).     
\label{eq:eqI2Iinference}
\end{equation}

\begin{figure}[htbp]
    \centering
    \begin{tabular}{cc}
    
        \begin{tikzpicture}[spy using outlines={circle,yellow,magnification=2.5,size=1.6cm, connect spies}]
            \node {\includegraphics[width=3.3cm]{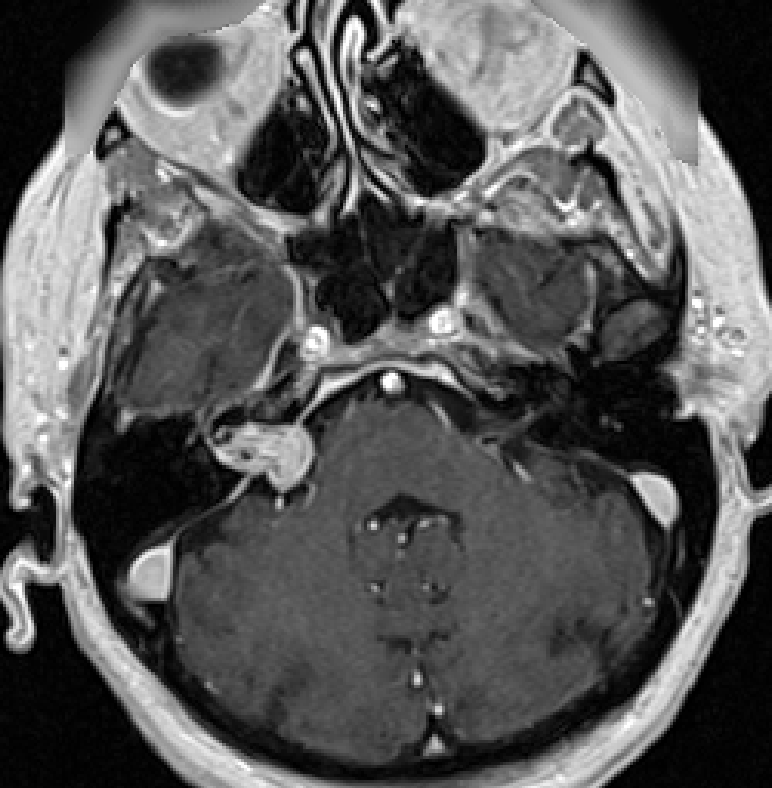}};
            \spy on (-0.50,-0.25) in node [left] at (1.5,0.8);
        \end{tikzpicture} &
        \begin{tikzpicture}[spy using outlines={circle,yellow,magnification=2.5,size=1.6cm, connect spies}]
            \node {\includegraphics[width=3.3cm]{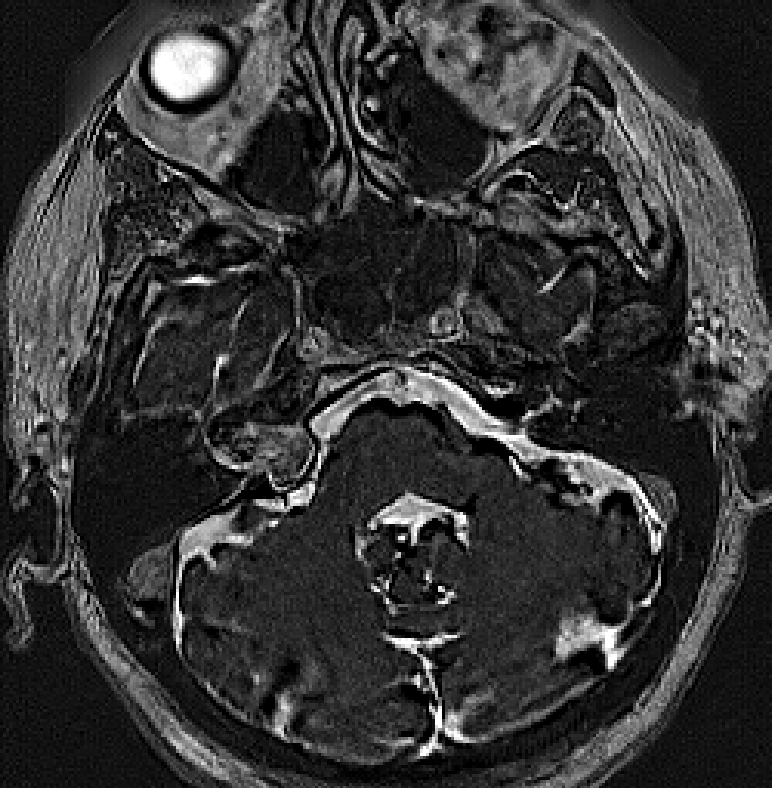}};
            \spy on (-0.50,-0.25) in node [left] at (1.5,0.8);
        \end{tikzpicture} \\
        (a) &
        (b) \\

        \begin{tikzpicture}[spy using outlines={circle,yellow,magnification=2.5,size=1.6cm, connect spies}]
            \node {\includegraphics[width=3.3cm]{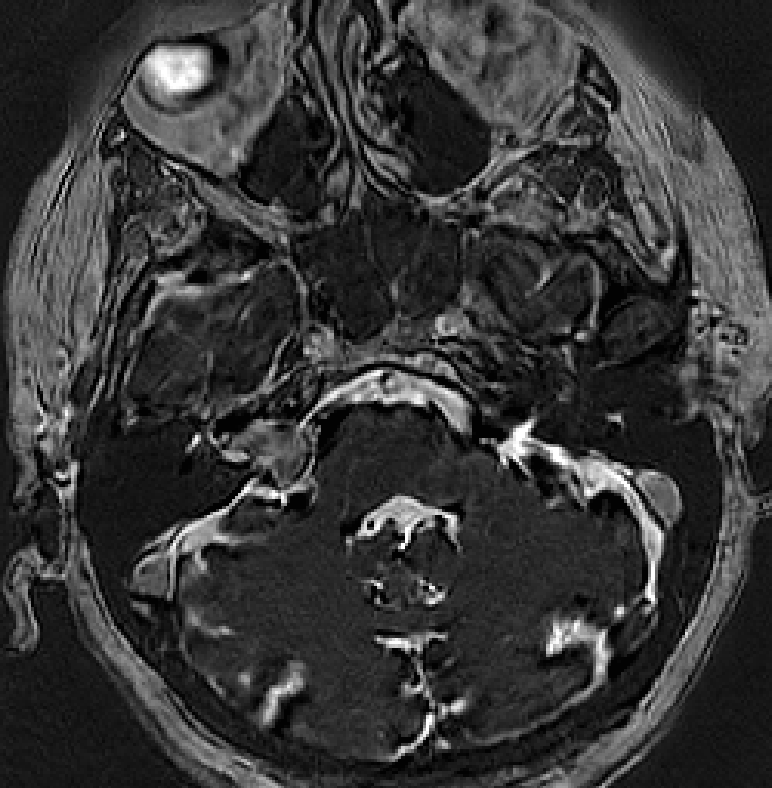}};
            \spy on (-0.50,-0.25) in node [left] at (1.5,0.8);
        \end{tikzpicture} &
        \begin{tikzpicture}[spy using outlines={circle,yellow,magnification=2.5,size=1.6cm, connect spies}]
            \node {\includegraphics[width=3.3cm]{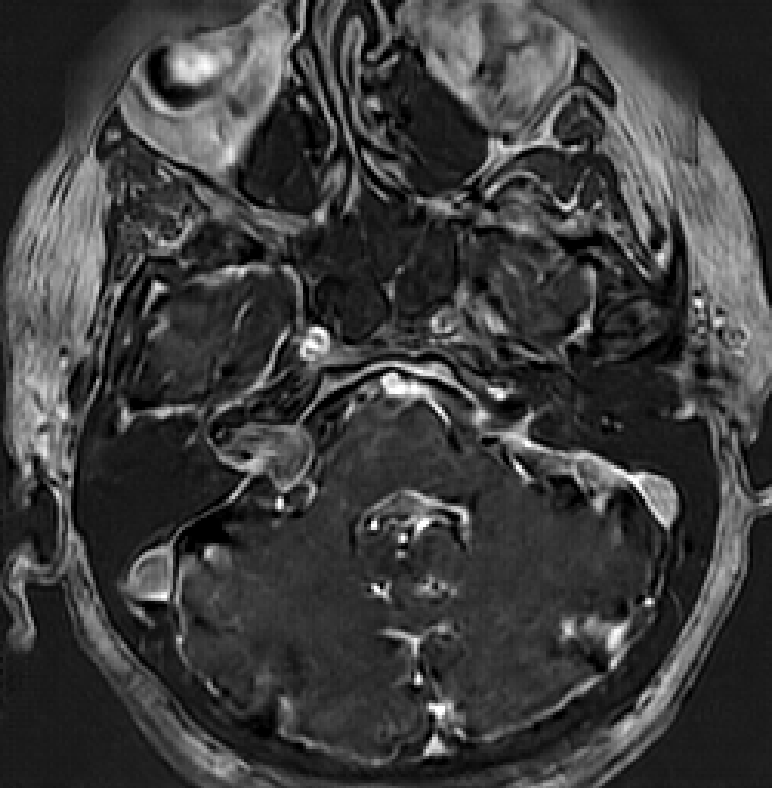}};
            \spy on (-0.50,-0.25) in node [left] at (1.5,0.8);
        \end{tikzpicture} \\
        (c) &
        (d) \\
    \end{tabular}
    \caption{Illustration of the instability of CycleGAN outputs in the ceT1$\rightarrow$hrT2 task of CrossMoDA 2022, affecting tumor appearance (close-up view). (a) original ceT1 image (b,c,d) synthetic hrT2 images for three CycleGAN runs with identical hyperparameters. } \label{cgfig}
\end{figure}

While the global appearance of generated  images $\tilde{X^T}$ is similar to $X_A^T$ (Fig.~\ref{cgfig}), critical limitations are identified in the translation. First, when analyzing features more locally (i.e. around the tumor), pseudo-images $\tilde{X^T}$ show altered VS features and weaker VS contrast compared to real $X^T$ images (Fig.~\ref{cgfig}, close-up). This is in line with previous observations that CycleGANs tend to ignore smaller scale features like tumors~\cite{cohen}. Second, CycleGAN outputs are highly variable, as demonstrated through retraining with identical hyperparameters (Fig.~\ref{cgfig}b-d). This critical aspect is often overlooked, resulting in unacceptable variations in tumor appearance depending on the CycleGAN run considered. As a consequence, these variations have a detrimental impact on downstream segmentation tasks. Lastly, since we used only one source center $A$, images from center ${B}$ cannot be mapped faithfully to pseudo-target images as their distribution is not spanned by the $X_S^A$ distribution. 

For these reasons, our objective is to increase the diversity of $\tilde{X^T}$ images encountered by the downstream segmentation model. Specifically, we aim to augment the pseudo-target images in order to expand the range of tumor appearances that the model is exposed to during training by: 1) correcting the local appearance of pseudo-target images produced by CycleGANs at the tumor level \textit{post-hoc} 2) compensating for the aleatoric nature of CycleGAN runs and 3) addressing potential domain gaps that may arise at test time due to the center effect within each modality. To this end, we propose to realistically blend altered versions of pseudo-target tumors using a deep generative model described in the next section.

\subsection{Generative Blending Augmentation}

We propose a new data augmentation technique for tumor segmentation based on a modified SinGAN model~\cite{Shaham2019} described on Fig.~\ref{singan}. It is a {2D} multi-scale generative model able to learn useful generative features from a unique image that has seldom been used in medical imaging ~\cite{micgs,micgs2022,thambawita2022singan}. We realistically blend naively contrast-altered tumor volumes in surrounding tissues using the learned SinGAN generators, a method we call Generative Blending Augmentation (GBA), summarized on Fig.~\ref{workflowpart1}. We focus in particular on tumor contrast alteration stemming from the visual observation that VS contrast values in CrossMoDA are highly variable across subjects, ranging from hypo- to hypersignal with respect to surrounding healthy tissues. Our hypothesis is thus that generating unseen VS contrasts may help combat the domain shift at test time. 

For each pseudo-target image in $\tilde{X^T}$ (with corresponding label from $Y^S$), a straightforward technique consists in multiplying the tumor intensity levels by an intensity scaling factor $\lambda$: 

\begin{equation}
\forall (x,y) \in (\tilde{X^T}, Y^S), x_\lambda =
\begin{cases}
    x & \text{outside the tumor,} \\
    \lambda \cdot x & \text{inside the tumor.} \\
\end{cases}
\label{eq:eq1a}
\end{equation}

A value $\lambda < 1$ thus artificially lowers the signal of the tumor with respect to the surrounding tissues (i.e. it makes it darker in gray scale), while a value $\lambda > 1$ increases it. While this is helpful to diversify the training set, such a naive change leads to unrealistic tumor appearance, especially at the tumor boundaries. To more realistically blend the altered tumors to the target image, we use the object harmonization ability of SinGAN. SinGAN, described on Fig.~\ref{singan}, uses a series of cascaded generators trained sequentially at multiple spatial scales for unconditional image generation (i.e. from pure noise) using a single training image. Here, we leverage the learned generators to transfer the style {of a real target 2D slice to alter and harmonize the tumor slices within $x_\lambda$.} 
More formally, a SinGAN network is composed of $K+1$ generators and discriminators $(G_k, D_k)_{0 \leq k \leq K}$ arranged in a coarse-to-fine fashion (Fig.~\ref{singan}). Each GAN pair learns the distribution of training image patches at a spatial scale $k \in (0,...,K)$, where $K$ refers to the coarsest scale and $0$ to the finest, original scale. Once a scaling factor between scales $r$ is fixed, the number of scales $K$ is automatically determined so as to achieve a $25 \times 25$ image at the coarsest scale and the original size at the finest.

We first train a 2D GAN at the coarsest level $(G_K, D_K)$ by considering only Gaussian noise as input (i.e. unconditional GAN setting). Once trained, its weights are fixed and the output is upsampled once to condition the training of the GAN of the scale above it $(G_{K-1}, D_{K-1})$. 

In order to achieve realistic blending of the altered tumor, we fix a spatial scale level $k^\ast$, i.e. the number of learned generators used to control the degree of blending (and thus harmonization quality). For every 2D pseudo-target slice {$x_\lambda(z)$} containing tumor tissues, we downsample it $k^\ast$ times by the scaling factor $r$ (arrow notation $\downarrow_{r^{k^\ast}}$). We then forward pass the image using into each specified generator $(G_k)_{0 \leq k \leq k^\ast}$ and upsample the output (notation $\uparrow_r$): 
\begin{equation}
\begin{aligned}
    G^{up}_k =& \uparrow_r \circ \; G_k \\
    \psi_{k^\ast}({x_\lambda(z)}) = (G_{0} \circ G^{up}_{1} &\circ ... \circ G^{up}_{k^\ast} \circ \downarrow_{r^{k^\ast}})({x_\lambda(z)}) \\
\end{aligned}
\label{eq:eq1b}
\end{equation}


{We then aggregate all $\psi_{k^\ast}({x_\lambda(z)})$ to reconstruct a 3D volume. To simplify notation, we denote this volume as $\psi_{k^\ast}({x_\lambda})$.} We finally merge  {the resulting volume with the unaltered pseudo-target slices to form the GBA-augmented image}. To further smoothen the transition between object and background, the binary mask $y$ of the tumor is dilated and convolved with a Gaussian filter in the axial plane to form a pixel weighting mask $\hat{y}$ between the augmented image and the original image: 
\begin{equation}
\psi_\text{GBA}(x_\lambda, \hat{y}, k^\ast) = \hat{y} \cdot \psi_{k^\ast}(x_\lambda) + (1-\hat{y}) \cdot x_\lambda.
\label{eq:eq1c}
\end{equation}






We also consider the more naive data augmentation strategy, hereafter mentioned as Naive tumor rescaling, where we removed the SinGAN blending step. In this case, instead of harmonizing the rescaled tumor, we simply paste it back on the pseudo-target image without blending. Doing so enables us to independently evaluate the tumor augmentation strategy and the blending step.

\subsection{Segmentation with self-training}

\begin{figure}[htbp]
\begin{center}
\includegraphics[width=0.9\linewidth]{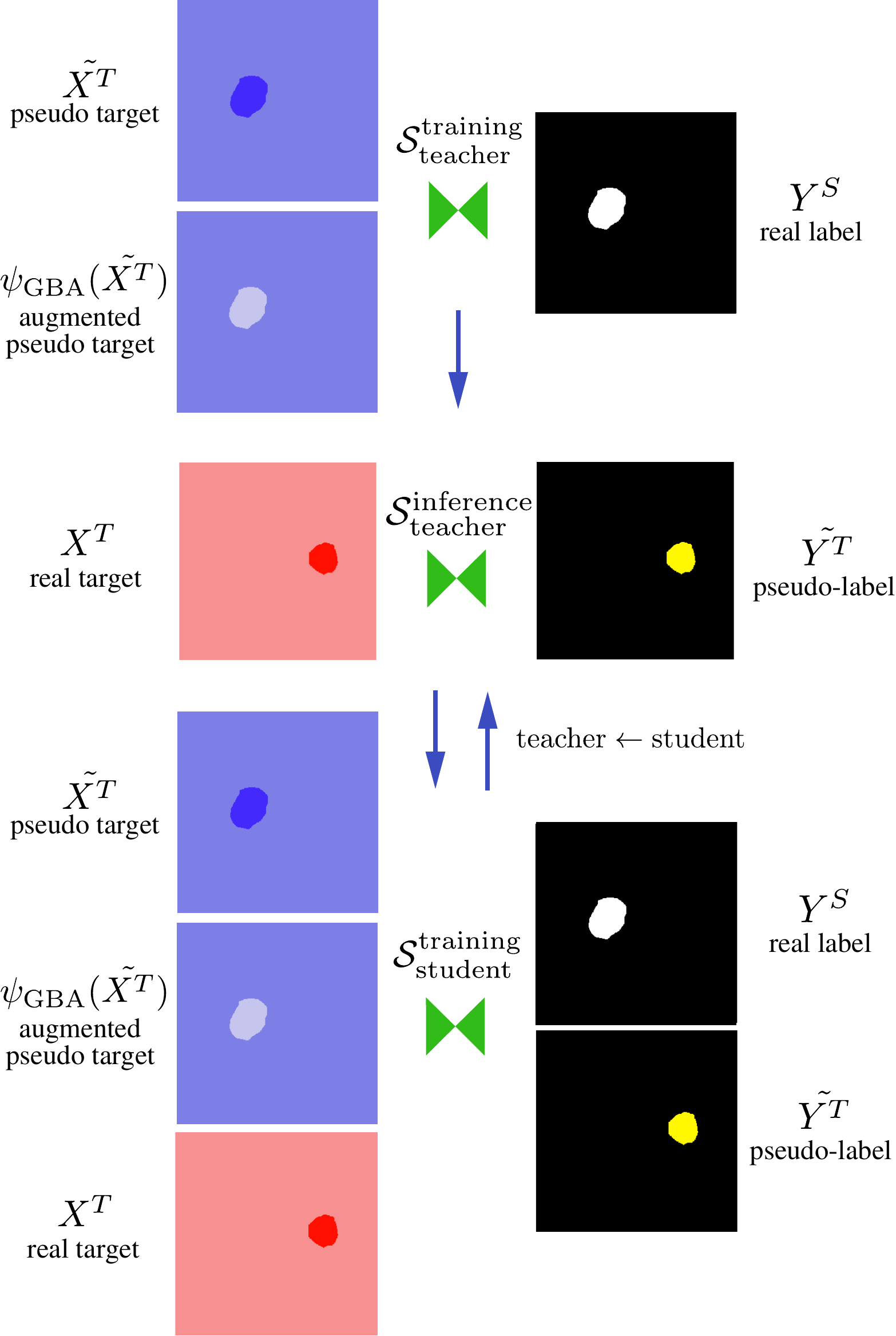}
\caption{Iterative segmentation and self-training with GBA. Red and blue thumbnails respectively stand for real and generated images, while yellow-colored ROI segmentation masks represent pseudo-labels. After a first segmentation pass with a \textit{teacher} segmentation model trained on pseudo-target images with GBA, we generate pseudo-labels using the unlabelled target set. Then, a \textit{student} model is retrained, enriched with target images and pseudo-labels. The \textit{student} replaces the \textit{teacher} model and this loop is repeated several times to enhance pseudo-labels quality iteratively.}\label{workflowpart2}
\end{center}
\end{figure}

Concurrently to the proposed augmentation techniques (tumor intensity rescaling and GBA blending refinement), we leverage an iterative self-training scheme for tumor segmentation, that was also proposed by the best performing method of the previous edition of CrossMoDA~\cite{xie2020self,cosmos}. This workflow is described on Fig.~\ref{workflowpart2}. We perform supervised training using both real labels aligned to the GBA-augmented pseudo-target images, but also pseudo-labels from real target images after a first segmentation forward pass, i.e. using a \textit{teacher}-\textit{student} training strategy. 

A first segmentation model $\mathcal{S}_\text{teacher}$ is trained on GBA-augmented I2I outputs $(\tilde{X^T} \cup \psi_\text{GBA}(\tilde{X^T}), Y^S)$. A forward pass of $\mathcal{S}_\text{teacher}$ is then performed on $X^T$ to produce pseudo-labels $\tilde{Y^T}$ (which are typically noisy due to the domain shift). The training data distribution is thus divided into  $(\psi_\text{GBA}(\tilde{X^T}), Y^S)$ with augmented pseudo-target images and real labels and $(X^T, \tilde{Y^T})$ with real images and pseudo-labels. Following~\cite{xie2020self}, we combine these sets for improved robustness. We repeat this loop and replace pseudo-labels $\tilde{Y^T}$ at each iteration, until no improvement on validation segmentation performance is observed, (measured by a variation in Dice lower than $0.3\%$). In practice, convergence is reached after $3$ iterations.

\section{Validation setup}

In this section, we describe the datasets, evaluation metrics, pre-processing and practical details used in our experiments that focused on the CrossMoDA 2022 segmentation challenge.

\subsection{Challenge dataset, metrics and ranking} \label{dataset}

\begin{figure}[H]
\begin{center}
\includegraphics[width=1.0\linewidth]{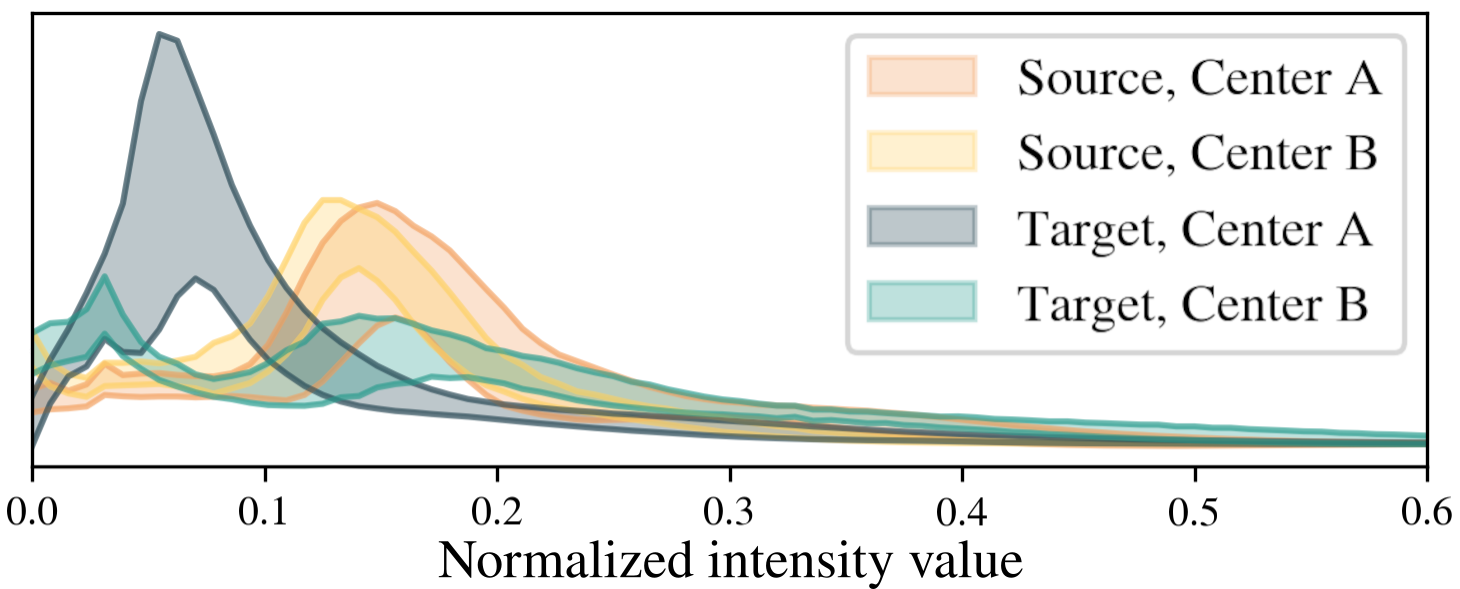}
\caption{Normalized intensity distributions of source and target image modalities per center. Background voxels were excluded and images were rescaled to the range $[0,1]$. Each band spans one standard deviation below and above the mean.}\label{histpercenter}
\end{center}
\end{figure}

The CrossMoDA 2022 training set is composed of $210$ labeled ceT1 images and $210$ unpaired, unlabelled hrT2 images, equally populated by two clinical centers: Center A (\textit{London}) and Center B (Tilburg) ~\cite{shapey2021segmentation}. Normalized intensity distributions are shown in Fig. ~\ref{histpercenter}, highlighting substantial intensity domain shifts depending on the center considered, especially among hrT2 images.

ceT1 images from Center A showed an in-plane resolution of $0.4 \times 0.4$ mm$^2$ and in-plane matrices of $512 \times 512$, with varying slice thickness ranging from $1.0$ to $1.5$ mm. hrT2 images showed an in-plane resolution of $0.5 \times 0.5$ mm and in-plane matrices of $384 \times 384$ or $448 \times 448$, with varying slice thickness ranging from $1.0$ to $1.5$ mm. Center B showed ceT1 in-plane resolution of $0.8 \times 0.8$ mm and in-plane matrices of $256 \times 256$, with slice thickness of $1.5$ mm, while hrT2 images had in-plane resolutions of $0.4 \times 0.4$ mm and in-plane matrices of $512 \times 512$, with slice thickness of $1.0$ mm. All VS segmentation masks were manually delineated on selected axial slices in consensus by a neurosurgeon and an oncologist on the ceT1 images using both ceT1 and hrT2 modalities as guidance. A 2D semi-automated segmentation software was then employed to produce 3D segmentation masks. No reference label was given for any target image.

\begin{figure*}[!t]
    \centering
    \begin{tabular}{ccc}
        \begin{subfigure}[b]{0.30\linewidth}
            \centering 
            \includegraphics[width=\textwidth]{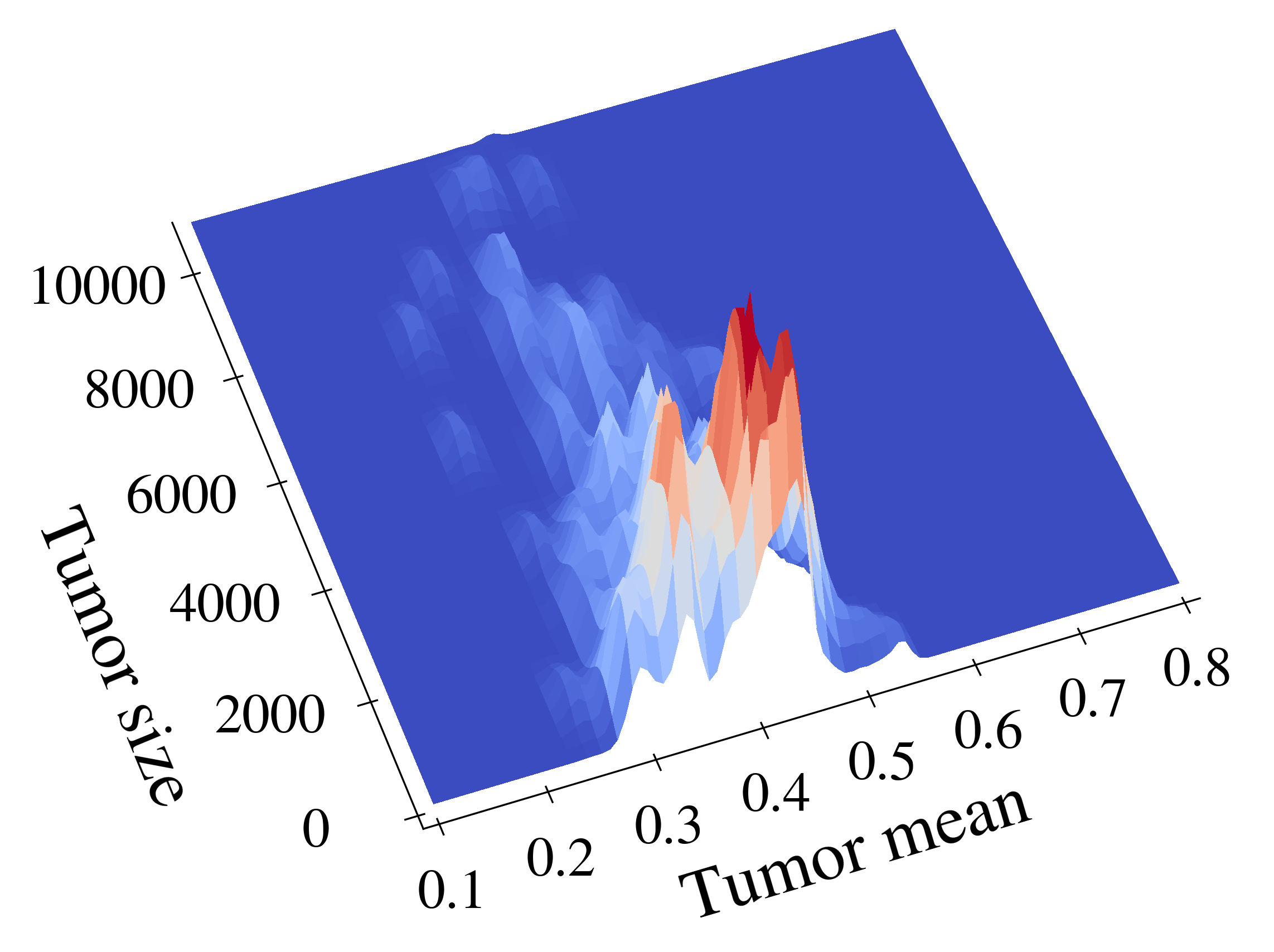}
            \caption{}
            \label{kde1}
        \end{subfigure} &
        \begin{subfigure}[b]{0.30\linewidth}
            \centering 
            \includegraphics[width=\textwidth]{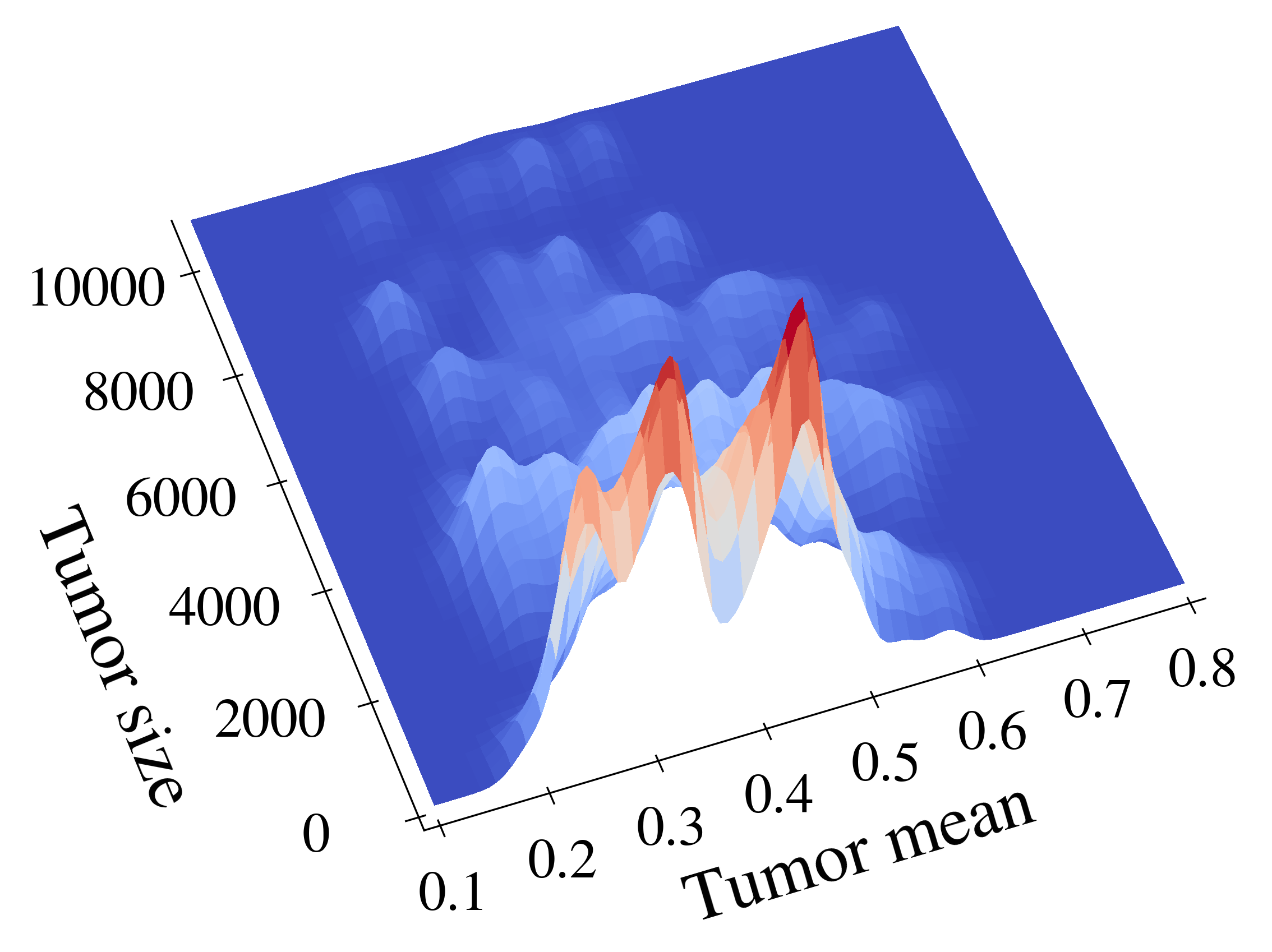}
            \caption{}
            \label{kde2}
        \end{subfigure} &
        \begin{subfigure}[b]{0.30\linewidth}
            \centering 
            \includegraphics[width=\textwidth]{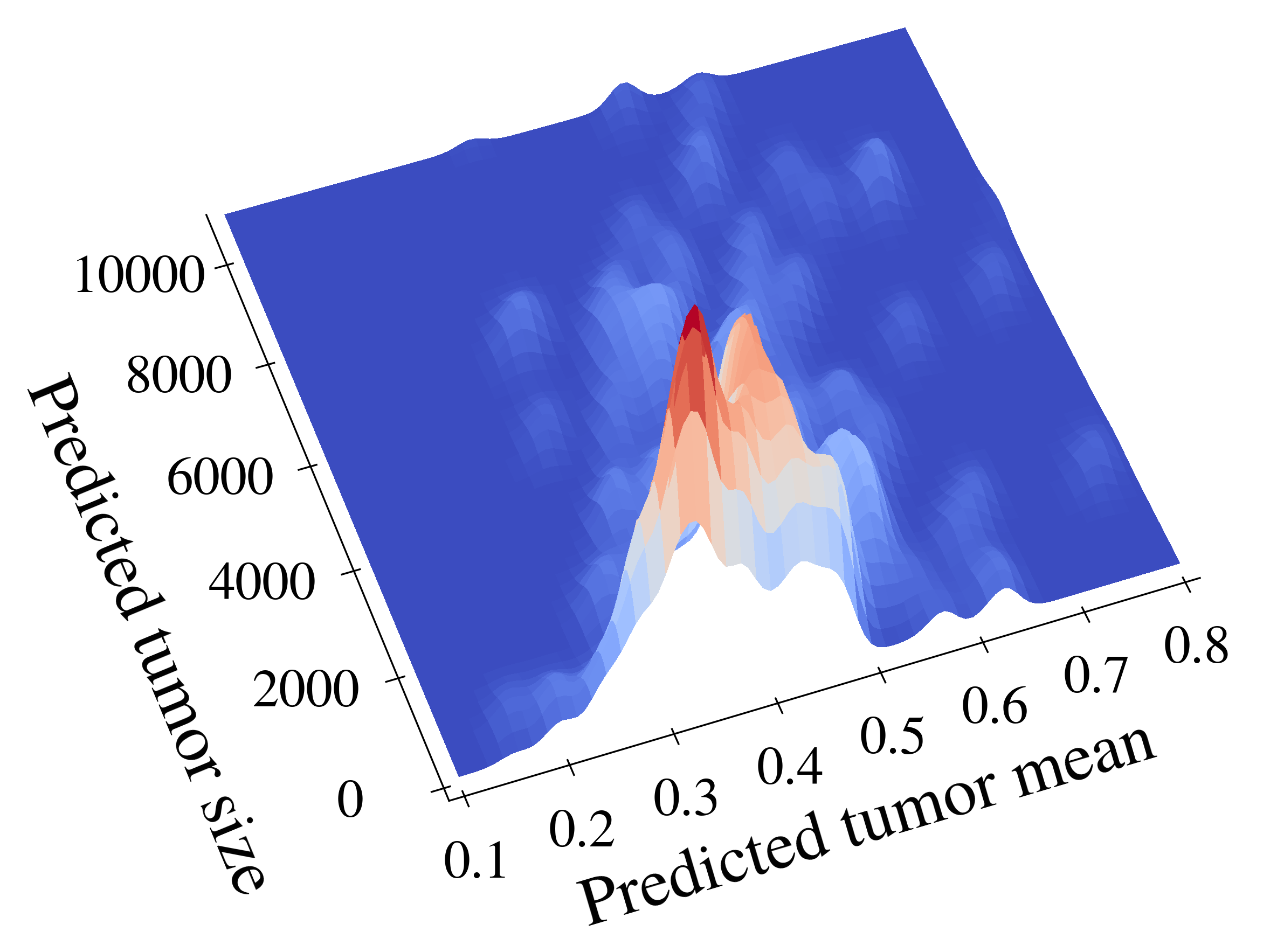}
            \caption{}
            \label{kde3}
        \end{subfigure} \\
    \end{tabular}
    \caption{2D kernel density estimations of mean tumor volume and normalized image intensity in the CrossMoDA challenge dataset (a) pseudo hrT2 images without augmentation, (b) GBA-augmented pseudo hrT2 images and (c) real target hrT2 images. The distribution of GBA-augmented images spans most of the target distribution.} \label{KDE}
\end{figure*}

A validation set was also accessible to the challengers, containing $32$ unannotated hrT2 images from each center ($64$ in total). To evaluate the method during and after the challenge without granting access to ground truth labels, validation was only possible online with one daily submission returning two metrics per patient: Dice similarity coefficient (DSC) and average symmetric surface distance (ASSD).

Another $270$ patients ($137$ from London, $133$ from Tilburg) remained private for testing and ranking purposes. The challenge evaluated segmentation performance on two organs, the VS and the main organ at risk cochlea. However, cochlear segmentation from ceT1 images in CrossMoDA is a less relevant clinical objective as cochlear masks were actually delineated on (hidden) hrT2 images, as it is the case in clinical practice, and artificially transposed back to ceT1 images~\cite{shapey2021segmentation}. The clinical challenge is thus the segmentation of VS from hrT2 images, which is the focus of this paper. 

To rank participants during the challenge, results were sorted in terms of Dice and ASSD per subject for both classes cochlea and VS. An average rank for all patients the classes was computed to rank challengers globally. We focus 
in this article on the VS segmentation task as our approach did not aim at improving cochlear segmentation.

\subsection{Implementation details}

\subsubsection{GBA and SinGAN parameters}

The rationale for GBA augmentation parameters was to span the distribution of tumor hrT2 appearances observed in the two centers $A$ and $B$, requiring to tune the contrast scaling values $\lambda$ and the number of augmentations towards this objective. To guide the choice of these parameters, we used pseudo-labels produced by the first pass of the segmentation stage to sample a surrogate ground truth distribution of tumor intensities in the real hrT2 set. Fig.~\ref{KDE} compares the distribution of tumor mean image intensity and size before (Fig.~\ref{KDE}a) and after GBA (Fig.~\ref{KDE}b) in light of this surrogate ground truth distribution (Fig.~\ref{KDE}c). Contrast values $\lambda$ were tuned to span similar ranges of intensity and volumes (Fig.~\ref{KDE}b). This empirical tuning led to two major effects: 
\begin{itemize}
    \item The proportion of large heterogeneous tumors in the training set was increased as the Tilburg set contained generally larger tumors than the London set. We augmented all VS from \textit{Tilburg} center ($\tilde{X_B^T}$) with a volume larger than $2000$ mm$^3$ and standard variation value higher than $0.10$ ($29$ patients in total). This led to considering $3$ different $\lambda$ values: $0.7$, $1.2$ and $1.5$. The SinGAN model was trained with $K=16$ total  scales. For each $\lambda$, we generated $2$ samples retaining $k^\ast=1$ and $k^\ast=3$ scales. In total, this led to $2\times 3=6$ augmentations per patient.
    
    \item The contrast of smaller tumours was reduced. Images from both centers ($\tilde{X^T}$) showing VS smaller than $300$ mm$^3$ ($19$ patients in total) were augmented with smaller $\lambda$ values: $0.6$, $0.8$ and $1.2$. Identical parameters ($K$ and $k^\ast$) were used,  resulting also in $6$ augmentations per patient. 
\end{itemize} 

We used the original SinGAN architecture consisting of generators $\lbrace G_k\rbrace$  made of $5$-layer convolutional nets with  $3\times 3$ conv-blocks equipped with batch normalization and followed by LeakyReLu activations at each layer. The discriminators $\lbrace D_k\rbrace$ are PatchGANs  \cite{li2016precomputed} with a receptive field of $11 \times 11$ pixels. Computational time being not a concern for the challenge, we trained $K=16$ scales, which took around $8$ hours to train on a GTX 1080Ti GPU.

\subsubsection{Image resampling}

We resampled all ceT1 and hrT2 images to an identical voxel spacing of $0.6 \times 0.6 \times 1.0$ mm$^3$. Then, we automatically derived the coordinates ($x,y$) of the center of the brain as the average location of voxels higher than the 75$^\text{th}$ intensity percentile, as proposed in \cite{choi}. For each patient, we then cropped a sub-volume  $256 \times 256 \times Z$ centered on $(x,y)$,  where $Z$ represents the number of axial slices. As previously mentioned, due to excessive domain shift between clinical centers, we used only image data from (\textit{London}, $A$) to stabilize CycleGAN training and achieve ceT1 to hrT2 translation. The cyclic loss weight was set to $\mu_{cyc}=10$. Due to a blurring effect induced by the CycleGAN inference pass, all pseudo-target images were deconvolved with an iterative Van Cittert  (VC) deconvolution algorithm\cite{vancittert} that showed to slightly improve performance on the validation set. We used a Gaussian point spread function of scale $1\times1\times2.5$ mm$^3$ for $15$ VC iterations. Before the last segmentation iteration, we resampled the real hrT2 images to a finer voxel spacing of $0.4 \times 0.4 \times 1.0$ mm$^3$. 


\subsubsection{Segmentation parameters}

We used a 5-fold ensemble 3D full resolution nnU-Net model~\cite{nnunet} and performed 3 additional iterations of self-training with pseudo-labels, i.e. we trained 20 folds in total. Validating each segmentation training through 5-fold cross validation and ensembling allows to get trustful pseudo-labels that can be employed during the following iteration. Misclassified voxels may indeed impact the training significantly. Default nnU-net augmentations were used in combination with GBA: random rotations, scaling, addition of Gaussian noise, Gaussian blurring, brightness change, global contrast change, aliasing, gamma correction and mirroring. We trained all models for $500$ epochs except for the last pass that we trained for $1000$ epochs. The largest connected component after softmax thresholding was taken as the VS segmentation mask.

\subsection{Experiments} \label{evaluation}


Alongside final challenge segmentation results, we performed a series of experiments to evaluate in more detail the interest of the proposed GBA and naive rescaling augmentation, combined with a self-training strategy for cross-modal segmentation. 

\begin{figure}[htbp]
    \centering
    \begin{tabular}{ccc}
        \begin{subfigure}[b]{0.28\linewidth}
        \begin{tcolorbox}[
      colframe=red,
      boxrule=1pt,
      sharp corners,
      notitle,
      boxsep=0pt,
      left=1pt,
      right=1pt,
      top=1pt,
      bottom=1pt,
      colback=white
    ]
            \centering 
            \includegraphics[width=\textwidth]{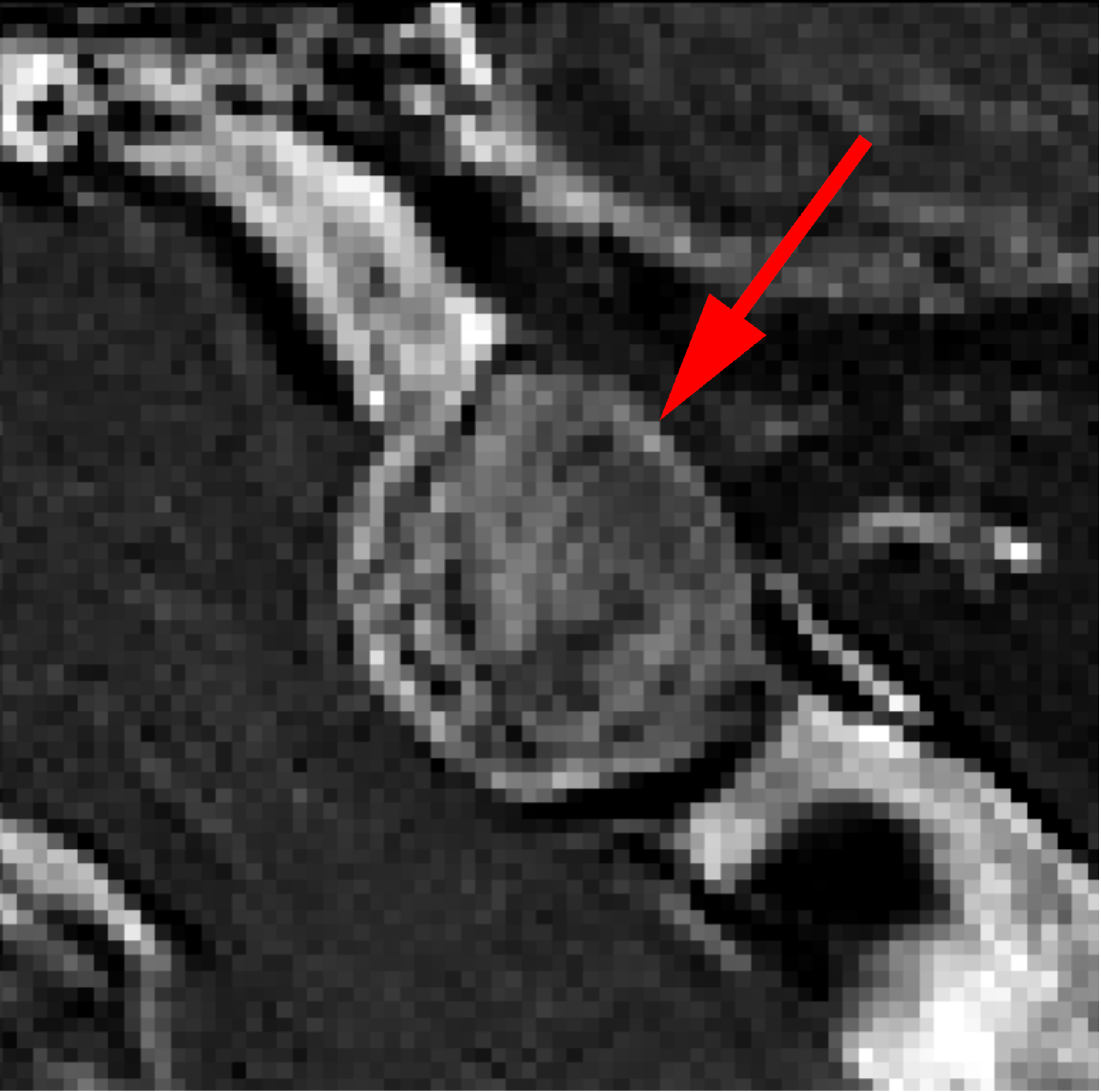} 
            \end{tcolorbox}
        \end{subfigure} &
        \begin{subfigure}[b]{0.28\linewidth}
            \centering 
            \includegraphics[width=\textwidth]{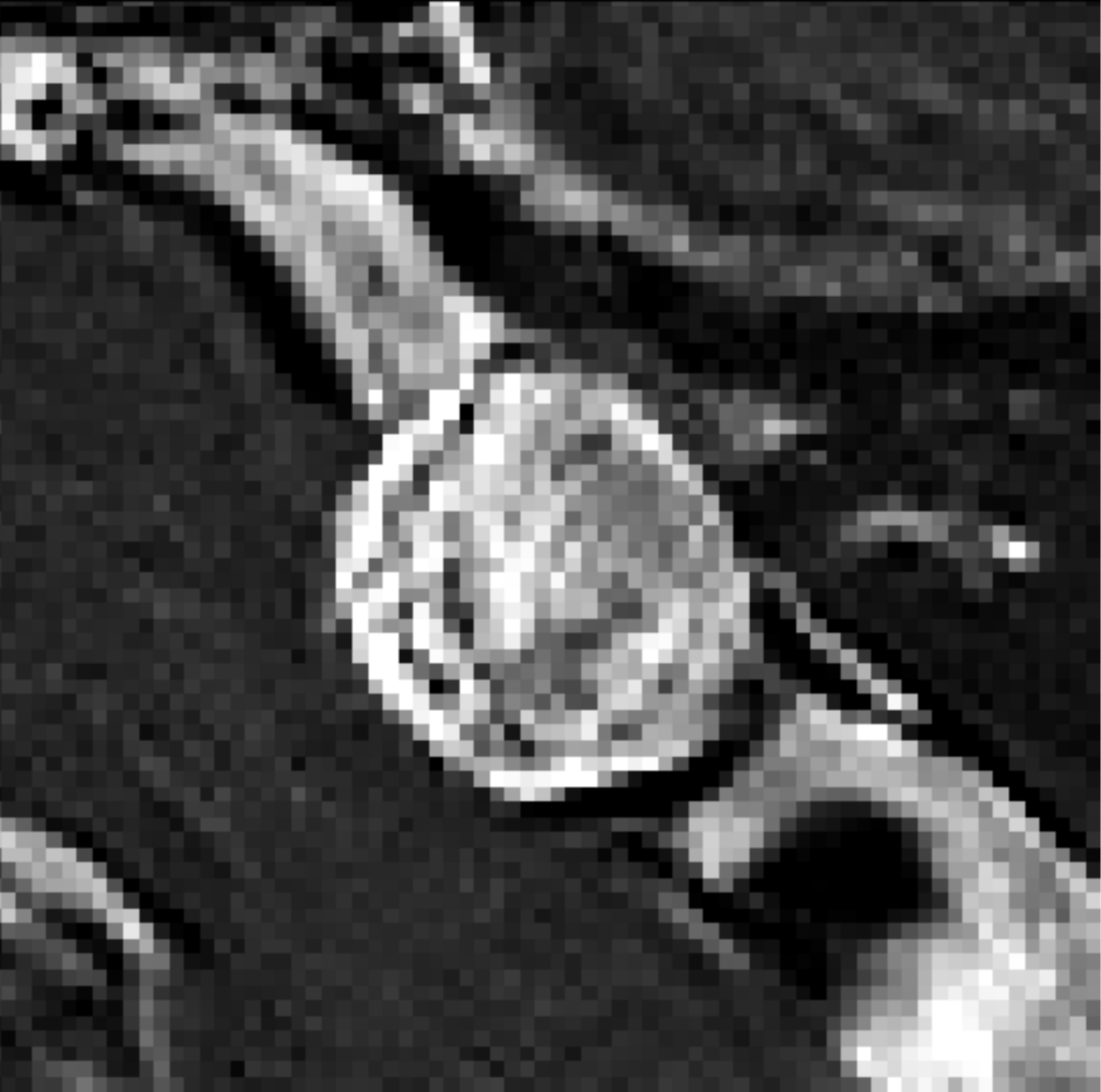} 
        \end{subfigure} &
        \begin{subfigure}[b]{0.28\linewidth}
            \centering 
            \includegraphics[width=\textwidth]{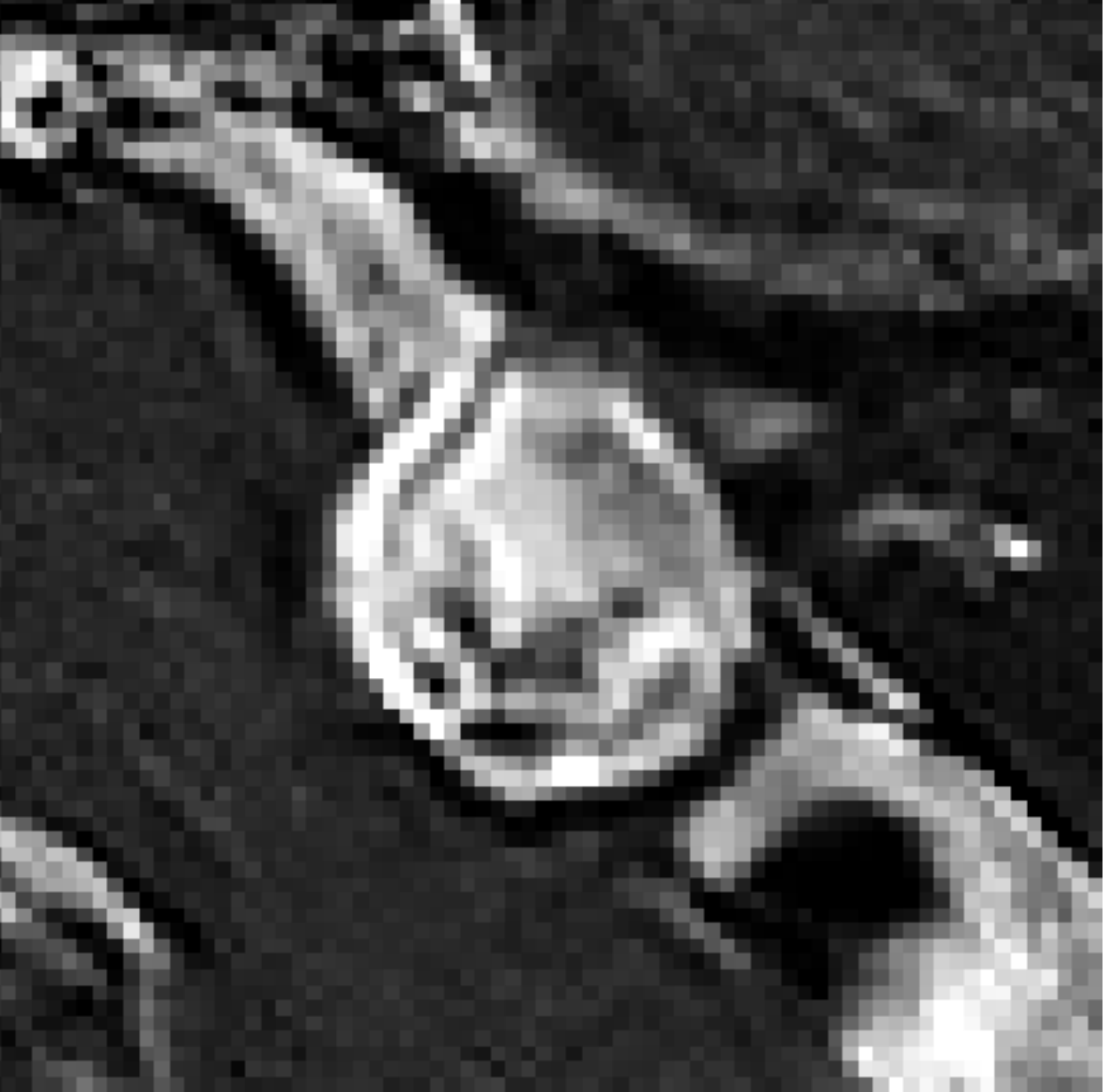} 
        \end{subfigure} \\
        
        \begin{subfigure}[b]{0.28\linewidth}
            \centering 
            \begin{tcolorbox}[
      colframe=red,
      boxrule=1pt,
      sharp corners,
      notitle,
      boxsep=0pt,
      left=1pt,
      right=1pt,
      top=1pt,
      bottom=1pt,
      colback=white
    ]
            \includegraphics[width=\textwidth]{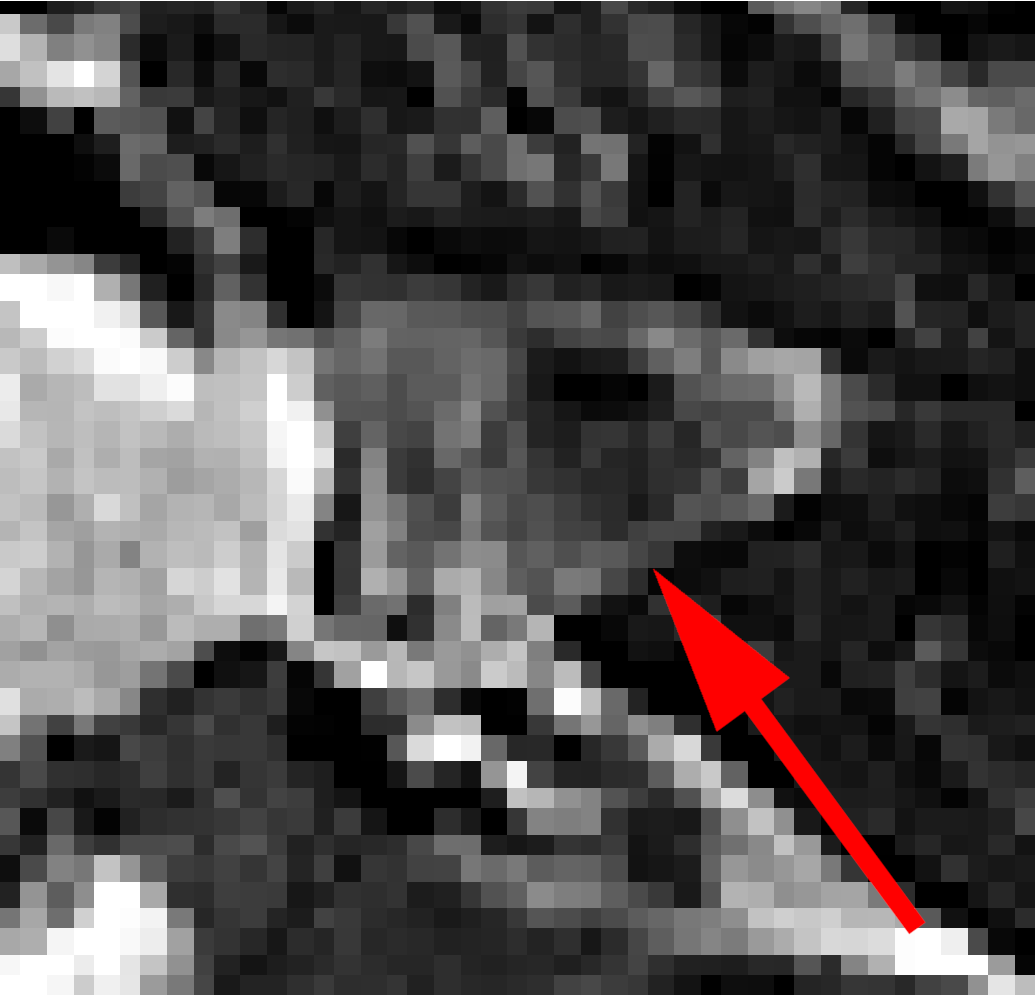}
            \caption{}
            \end{tcolorbox}
            \label{s}
        \end{subfigure} &
        \begin{subfigure}[b]{0.28\linewidth}
            \centering 
            \includegraphics[width=\textwidth]{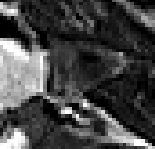}
            \caption{}
            \label{t}
        \end{subfigure} &
        \begin{subfigure}[b]{0.28\linewidth}
            \centering 
            \includegraphics[width=\textwidth]{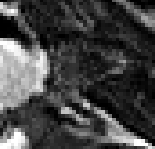}
            \caption{}
            \label{u}
        \end{subfigure}
    \end{tabular}
    \caption{Comparing GBA augmentation to naive tumor intensity rescaling in two patients of the validation set. (a) pseudo-target image (b) Naive tumor rescaling (c) GBA. } \label{augfig}
\end{figure}

We evaluated the ability of naive contrast alteration and GBA at correcting potential errors and training instabilities of the CycleGAN stage. We also analyzed the effect of the SinGAN tumor blending over naive rescaling. Finally, we evaluated the evolution of segmentation results over self-training iterations.

To this end, we submitted a variety of intermediate segmentation results to the validation leaderboard. First, we retrained CycleGAN models with identical parameters to measure the intrinsic instability of CycleGAN training and its impact on downstream segmentation. To study the influence of the center effect, we submitted results obtained using CycleGAN models trained on 1) \textit{London} only $(X_A^S,X_A^T)$ 2) \textit{Tilburg} only $(X_B^S,X_B^T)$ or 3) both centers $(X^S,X^T)$.  For all experiments, segmentation training was performed with a $5$ fold cross-validation nnU-Net ensembling, following nnU-Net practice. Erroneous translations that led to very poor segmentation performances after the first segmentation stage with pseudo-labels (i.e. average Dice score $\leq 0.4$) were considered as failure cases and did not undergo iterative self-training (to avoid useless energy consumption). In the remaining runs, we followed the evolution of segmentation scores along the self-training procedure. 

\section{Results}

\subsection{Qualitative analysis}
Fig.~\ref{augfig} compares the effect of naive augmentation and GBA on two example cases of the validation set with $\lambda > 1$ (higher tumor intensity) and $\lambda < 1 $ (lower tumor intensity). Compared to naive augmentation, GBA added local pattern variations inside the tumor and induced a subtle blur effect in the first subject (top). Interestingly, this blur was not reproduced in the second subject (bottom), where GBA produced a sharper output that illustrates its capacity at producing a nonlinear blending response. Naive rescaling results seemed less realistic with more artifacts around the VS boundaries in these two cases.

\begin{figure}[htbp]
    \centering
    \begin{tabular}{ccc}
         \begin{subfigure}[b]{0.29\linewidth}
            \centering 
            \includegraphics[width=\textwidth]{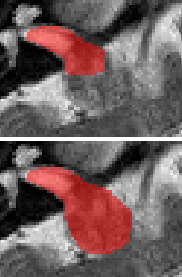}
        \end{subfigure} &
        
        \begin{subfigure}[b]{0.29\linewidth}
            \centering 
            \includegraphics[width=\textwidth]{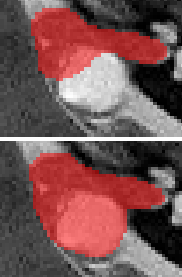}
        \end{subfigure} &
        
        \begin{subfigure}[b]{0.29\linewidth}
            \centering 
            \includegraphics[width=\textwidth]{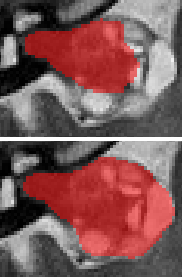}
        \end{subfigure} \\

    \end{tabular}
    \caption{Qualitative VS segmentation results for three patients of the validation set after one segmentation stage (training on pseudo-
target images and real labels only).Top: without GBA. Bottom: with GBA. \label{seg3D}}
\end{figure}

\begin{table*}[ht]
\begin{center}
\begin{tabular}{ll||c|c||c|c||c|cc}
        & &  \multicolumn{2}{c||}{nnU-Net augmentations}   &    \multicolumn{2}{c||}{+ Naive rescaling}  &  \multicolumn{2}{c}{+ GBA}   \\
       {\textbf{CycleGAN center}}  & {Run}& Dice $\uparrow$  & ASSD $\downarrow$ & Dice $\uparrow$  & ASSD $\downarrow$ & Dice $\uparrow$  & ASSD $\downarrow$    \\\hline
{London} & { \#1} & $0.679 \pm 0.341$  & $7.853 \pm 16.750$ & $0.789 \pm 0.241$ & $3.267 \pm 10.738$ & $\mathbf{0.832 \pm 0.159}$ & $\mathbf{1.315 \pm 6.235}$ \\
& { \#2} & $0.668 \pm 0.280$  & $3.653 \pm 10.857$ & $0.647 \pm 0.293$ & $2.185 \pm 6.341$ & $\mathbf{0.752 \pm 0.202}$ & $\mathbf{1.998 \pm 4.196}$ \\ 
& { \#3} & $0.711 \pm 0.289$  & $2.856 \pm 8.662$ & $0.812 \pm 0.146$ & $0.667 \pm 0.887$ & $\mathbf{0.816 \pm 0.144}$ & $\mathbf{0.638 \pm 0.850}$ \\ \hline 

{Tilburg} & { \#1} & $0.290 \pm 0.338$ & $21.060 \pm 34.855$ & N/A & N/A & N/A & N/A  \\ 
&  { \#2} & $0.284 \pm 0.370$ & $38.803 \pm 44.943$ & N/A & N/A & N/A & N/A \\ 
& { \#3} & $0.312 \pm 0.321$ & $18.989 \pm 33.111$ & N/A & N/A & N/A & N/A \\ \hline 
 
{London+ Tilburg} & { \#1}  & $0.243 \pm 0.354$  & $27.559 \pm 22.062$ & N/A & N/A & N/A & N/A  \\
&  { \#2} & $0.743 \pm 0.233$  & $2.116 \pm 7.419$ & $\mathbf{0.811 \pm 0.150}$ & $\mathbf{0.641 \pm 0.617}$ &  $0.800 \pm 0.173$ & $1.688 \pm 8.628$ \\
& { \#3} & $0.474 \pm 0.338$  & $7.714 \pm 15.357$ & $0.594 \pm 0.304$ & $\mathbf{2.743 \pm 7.347}$ & $\mathbf{0.612 \pm 0.304}$ & $4.060 \pm 10.349$ \\
\end{tabular}
\caption{Segmentation results on the validation leaderboard depending on the CycleGAN run considered, the center used as training data and the use of naive tumor rescaling or GBA. N/A indicates the experiment was not run due to clear failure of the baseline translation (DSC$\leq0.4$ without GBA). Conventional augmentations from nnU-Net were used in all experiments.}\label{scores_i2i}
\end{center}
\end{table*}

\begin{table*}[htbp]
\begin{center}
\begin{tabular}{ll||c|c||c|c||c|cc}
        & &  \multicolumn{2}{c||}{nnU-Net augmentations}   &    \multicolumn{2}{c||}{+ Naive rescaling}  &  \multicolumn{2}{c}{+ GBA}   \\
       {\textbf{Tumor set}}  & & Dice $\uparrow$  & ASSD $\downarrow$ & Dice $\uparrow$  & ASSD $\downarrow$ & Dice $\uparrow$  & ASSD $\downarrow$    \\ \hline
{All tumors} & {Real labels only}   & $0.774 \pm 0.220$  & $2.673 \pm 9.795$ & $0.839 \pm 0.126$ & $0.555 \pm 0.586$ & $\mathbf{0.847 \pm 0.114}$  & $\mathbf{0.514 \pm 0.510}$  \\ 
& {Self iteration 1}   & $0.780 \pm 0.219$  & $3.647 \pm 14.086$ & $0.859 \pm 0.067$ & $0.459 \pm 0.192$ & $\mathbf{0.862 \pm 0.065}$  & $\mathbf{0.444 \pm 0.182}$  \\ 
&  {Self iteration 2 }    & $0.771 \pm 0.237$  & $4.711 \pm 16.324$ & $0.860 \pm 0.066$ & $0.455 \pm 0.189$ & $\mathbf{0.865 \pm 0.062} $  & $\mathbf{0.438 \pm 0.184}$  \\ 
&   {Self iteration 3}    & $0.823 \pm 0.132$  & $0.598 \pm 0.541$ & $0.864 \pm 0.062$ & $0.447 \pm 0.186$ & $\mathbf{0.868 \pm 0.061}$  & $\mathbf{0.431 \pm 0.177}$ \\ \hline 
{Small tumors} & {Real labels only}   & $0.763 \pm 0.100$  & $0.441 \pm 0.192$ & $0.762 \pm 0.112$ & $0.501 \pm 0.337$ & $\mathbf{0.781 \pm 0.080}$  & $\mathbf{0.400 \pm 0.144}$  \\ 
& {Self iteration 3}  & $0.748 \pm 0.141$  & $0.482 \pm 0.303$ & $0.801 \pm 0.071$ & $0.366 \pm 0.123$ & $\mathbf{0.806 \pm 0.068}$  & $\mathbf{0.349 \pm 0.114}$  \\ \hline 
{Large tumors} &  {Real labels only}    & $0.786 \pm 0.174$  & $1.096 \pm 0.798$ & $0.890 \pm 0.043$ & $0.585 \pm 0.232$ & $\mathbf{0.893 \pm 0.038} $  & $\mathbf{0.571 \pm 0.198}$  \\ 
& {Self iteration 3}    & $0.802 \pm 0.183$  & $0.991 \pm 0.867$ & $0.901 \pm 0.037$ & $0.537 \pm 0.206$ & $\mathbf{0.905 \pm 0.035}$  & $\mathbf{0.513 \pm 0.192}$  \\ 
\end{tabular}
\caption{Segmentation results on the entire validation set along iterations of self-training. Due to variations in performance with tumor volume, results on two on the $15$ smallest and $15$ largest tumors ($<720$mm$^3$ or $>5400$mm$^3$) are also shown. Conventional augmentations from nnU-Net were used in all experiments.}\label{scores_aug}
\end{center}
\end{table*}

Fig.~\ref{seg3D} shows representative 2D segmentation results achieved with or without GBA in three \textit{Tilburg} patients after the first segmentation stage (i.e. only pseudo-target images and real labels were used as training data). No segmentation label was provided for any target image, so no ground truth is superimposed. In these examples, when not using GBA, important portions of the tumor volume were lost, contrary to GBA-based segmentation that better captured the full extent of the VS.

\subsection{CycleGAN variability assessment}

Tab.~\ref{scores_i2i} summarizes the segmentation scores achieved after the first segmentation stage (i.e. pseudo-target images and real labels only) using various CycleGAN training data configurations and retrainings with identical parameters. 

Using only \textit{London} data ($X_A$) yielded more stable results across CycleGAN retrainings, with e.g. a minimum average DSC of $0.668\pm0.280$. On the other hand, training using \textit{Tilburg} data only ($X_B$) led to erroneous cross-modal translations that in turn produced failed segmentations (DSC$\simeq0.3$, ASSD$\simeq20$ mm). As mentioned in Sec.~\ref{evaluation}, we did not pursue iterative segmentation when Dice scores were lower than $0.4$, as we considered this a failure case for the translation (marked N/A in the table). When combining \textit{London} and \textit{Tilburg} centers, segmentation results were not stable across multiple CycleGAN trainings, with large variations in terms of performance between retrainings (e.g. DSC=$0.243\pm0.354$ for {first} run, DSC=$0.743\pm0.233$ for {second} run). This justifies \textit{a posteriori} our choice of training the CycleGAN models on \textit{London} data only due to excess domain shift between centers. In each setting, adding GBA consistently improved metrics DSC and ASSD compared to nnU-Net augmentations only. The use of naive rescaling led to moderate improvements regarding these metrics (despite a slight drop of $-2.1\%$ in Dice score for {second} run {using \textit{London} data}). On the other hand, GBA results were more stable than naive rescaling. As expected, due to excess domain shift between centers, pooling the two centers led to very unstable results depending on the run considered.  The highest average DSC was achieved for the {first} GBA run {using \textit{London} data} with DSC=$0.832\pm0.159$, ASSD=$1.315\pm6.235$, and best ASSD for the {third} GBA run {using \textit{London} data} with ASSD=$0.638\pm0.850$, DSC=$0.816\pm0.144$.





\subsection{Iterative segmentation with GBA}

\begin{figure}[htbp]
\begin{center}
\includegraphics[width=1.0\linewidth]{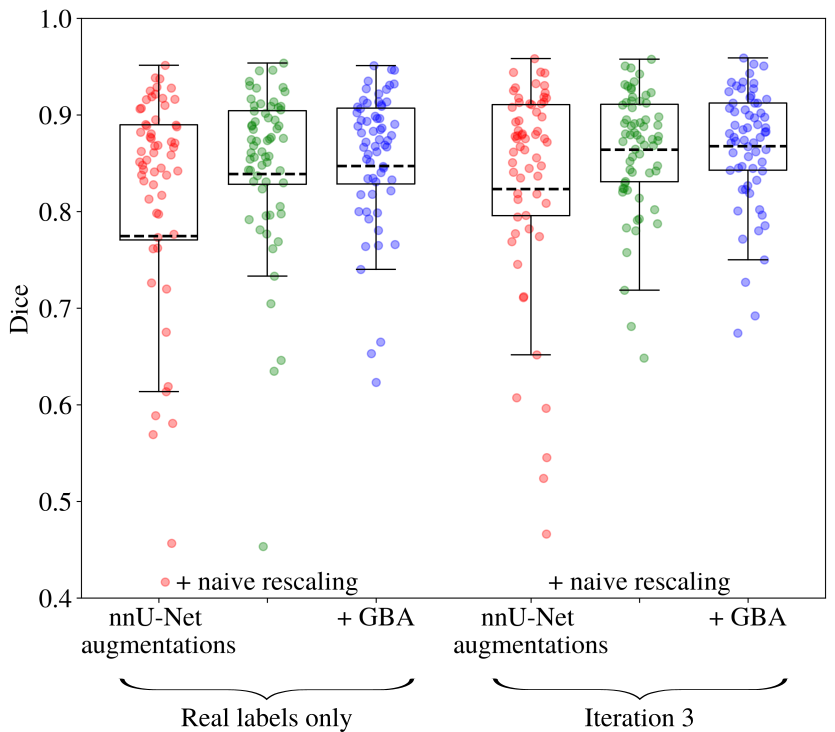}
\caption{Segmentation performances in the validation set considering only real labels of after iterative self-training. Three augmentation settings are considered : nnU-net augmentations only, or with addition of naive rescaling or GBA. Dashed lines show mean values. The number of cases with DSC $<0.4$ were respectively $3$, $1$, $1$, $2$, $0$ and $0$. These experiments correspond to the first and last rows of Tab.~\ref{scores_aug}.}\label{boxplotttt}
\end{center}
\end{figure}

Tab.~\ref{scores_aug} evaluates the improvement of segmentation scores across iterations of self-training with pseudo-labels in three augmentation settings 1) nnU-Net augmentations 2)  nnU-Net augmentations + naive tumor intensity rescaling and 3) nnU-Net augmentations + GBA. To further underline the behaviour of the various training strategy depending on the target tumor volume, we also present separated average results for small and large tumors. In the first segmentation pass (only pseudo-target images and real labels as training), both naive augmentations (DSC=$0.839\pm0.126$, p-value $=0.044$) and especially GBA (DSC=$0.847\pm0.114$, p-value $=0.021$) led to a sharp increase of segmentation performances over the nnU-Net baseline (DSC=$0.774\pm0.220$). 
Results were further improved with iterative self-training for 3 iterations for all methods, with a steadier increase for GBA and naive rescaling. GBA consistently achieved results superior to naive tumor rescaling. For instance, Dice overlaps were improved by $+0.8\%$ after the first segmentation step and $+0.4\%$ DSC after the last iteration of self-training for a maximum DSC=$0.868\pm0.061$ and minimum ASSD=$0.431\pm0.177$ mm. These differences between naive rescaling and GBA were not statistically significant according to a paired t-test (p-value $=0.7$). We observed similar trends when considering separately large and small tumors, with more than $10\%$ of Dice improvement for large tumors when employing naive rescaling or GBA with ground truth labels only. For small tumors, naive rescaling affected performance only slightly after the first self training iteration but improved more considerably thereafter, with an improvement of around $5\%$ of Dice for naive rescaling and $6\%$ for GBA.

Fig.~\ref{boxplotttt} compares the three augmentation settings, on the entire validation set, before the first and after the last iteration, using a boxplot visualization. With only nnU-Net augmentations, the number of badly segmented cases with DSC$<0.6$ was higher, regardless of the segmentation training considered. On the other hand a single segmentation learning based on GBA without any iterations performed better than after iterative self training without GBA.

\subsection{Final challenge test set}

Our team (LaTIM) achieved the first rank on the final leaderboard for the VS segmentation task (average VS rank score of $3.661$, mean DSC=$0.859\pm0.066$, mean ASSD=$0.459\pm0.252$) followed by team MAI-LAB (average VS rank score $3.862$, mean DSC=$0.852\pm0.094$, mean ASSD=$0.455\pm0.215$) and team ne2e (average VS rank score $3.948$, mean DSC=$0.856\pm0.064$, mean ASSD=$0.515\pm0.469$). The quantitative metrics associated to our method are summarized in Tab.~\ref{testingsettab}. Our worst result was DSC$\geq0.5$, ASSD$\leq2$ mm, which indicates that the proposed approach was able to identify the VS location at least partially for all subjects.

\begin{table}[ht]
\centering
\begin{tabular}{c|c|c}
LaTIM & Dice $\uparrow$ & ASSD (mm) $\downarrow$\\ \hline \hline
Mean & $0.859\pm0.066$ & $0.459\pm0.252$ \\ \hline
Worst & $0.502$ & $1.945$ \\ \hline
1st perc. & $0.637$ & $1.573$ \\ \hline
5$^\text{th}$ perc. & $0.747$ & $0.934$ \\ \hline
25$^\text{th}$ perc. & $0.831$ & $0.530$ \\ \hline
50$^\text{th}$ perc. & $0.868$ & $0.417$ \\ \hline
75$^\text{th}$ perc. & $0.905$ & $0.293$ \\ \hline
95$^\text{th}$ perc. & $0.937$ & $0.209$ \\ \hline
99$^\text{th}$ perc. & $0.947$ & $0.178$ \\ \hline
Best & $0.962$ & $0.165$ \\
\end{tabular}
\caption{CrossMoDA test set results (270 patients) for our team (LaTIM) decomposed into quantiles.}\label{testingsettab}
\end{table}

\section{Discussion}

The main objective of the present work was to study the potential interest of a new form of data augmentation, generative blending augmentation, to improve domain generalization for image segmentation. The underlying rationale behind GBA is to expose the network to a wide range of synthetic, realistically blended tumor contrasts to help tackle domain shifts during deployment. This approach is especially useful in scenarios where the distribution of tumor appearances during training may not cover the various deployment conditions due to factors such as sample selection bias or low data regimes. 

We positioned our work in the context of cross-modal segmentation as part of the 2022 CrossMoDA challenge. In this context, conventional cross-modal methods based on domain translation and segmentation are typically hard to train due to the presence of domain shifts within the source and target domains themselves (as they are composed of multiple clinical centers). As such, they cannot be considered as homogeneous domains from which meaningful translation can be learned using conventional I2I translation models like 2D CycleGAN. A potential major explanation for performance differences between centers can be the heterogeneity in image resolution, which  led to a reduced number of axial 2D slices for center B (Tilburg, 14000 slices) compared to center A (London, 30000 slices), as well as substantial intensity shifts between centers.

To overcome this challenge, we proposed a solution that involves restricting the domain translation process to a specific sub-domain (here $A$ for \textit{London} center), where a satisfying mapping could be achieved. We then performed appropriate augmentation based on GBA to expose the real downstream segmentation network to a diverse range of tumor appearances to be expected in the second domain. This approach can be adapted to increased number of centers through additional, center-specific augmentations.
Naturally, other improvements could have been considered, such as developing more complex neural network architectures that could tackle the domain shift and/or segmentation in a more integrative (e.g. end-to-end) fashion, or exposing the I2I network itself to similar data augmentations. Other high-performing teams in the challenge have indeed resorted to more complex I2I models (based on e.g. CUT~\cite{park2020contrastive} or NICE-GAN~\cite{chen2020reusing} or with the addition of segmenters inside the domain translation task~\cite{huo2018synseg}). Being a generic augmentation method, GBA may also be combined in principle to these approaches or to more recent diffusion-based I2I models~\cite{ozbey2022unsupervised}. However, our objective was to explore the interest of GBA independently. From this standpoint, the proposed tumor contrast alteration mechanism led to significant improvements in segmentation performances when used in combination to conventional, well-established data augmentation methods used in baseline models such as nnU-Net. 

While the fact that naive tumor intensity rescaling can impact segmentation performance by such a margin is an interesting and original result in itself, further blending the altered tumors to the target image using a cascade of SinGAN generators consistently improved downstream segmentation performances, albeit not significantly according to a paired t-test on the validation set (N=64). Due to its nonlinear harmonization capability, SinGAN blending allowed to harmonize the scaled tumors and better diversify their local  appearances over simple scaling (e.g. through local pattern variations, addition of realistic blur or sharpness alteration). We showed a clear advantage of this SinGAN blending step over naive tumor rescaling, especially in the early stages of iterative self training. This advantage remained but reduced along the iterations of self training. As SinGAN requires only a single MRI slice for training, this also makes it useful in a variety of training data regimes, with limited computational cost compared to other generative augmentation models trained on larger datasets.

We have submitted a number of experiments to the validation leaderboard to confirm that GBA was an essential ingredient in the success of our approach. Due to the limited number of submissions (one daily), many parameters were determined empirically to roughly optimize scores on the validation leaderboard. Fine-tuning them would have likely led to better results. For instance, we selected our contrast scaling values $\lambda$  through the inspection of the distribution of tumor intensity in the target set using pseudo-labels after the first segmentation stage, which is sub-optimal. Finding a better heuristic to adapt these parameters on the fly along the self-training procedure could  have likely allowed to achieve superior results.
 
While a self-supervised iterative strategy with pseudo-labels significantly improved segmentation results, it is not without fundamental limitations. First, the computational cost is strongly increased when leveraging iterations using self-training. Using GBA, we substantially improved performance even before self training, demonstrating the complementary nature of the two strategies. Second, erroneous predictions (i.e. party or totally missed tumors) that are then re-used downstream may be highly detrimental to the final model performance. This can explain the instability of self-training when only nnU-Net augmentations were used (Tab.~\ref{scores_aug}). Both naive tumor scaling and GBA yielded  statistically significant increase of segmentation performances after the first segmentation pass compared to nnU-Net augmentations. Even though performances between GBA and naive tumor rescaling were not statistically significant on the validation set constituted of only 64 patients, we posit that exposing the network to a diversity of more realistic tumor appearances at each iteration  with GBA is likely to have been key to the success of our approach. 

\section{Conclusion}

In this paper, we have presented a method for unsupervised cross-modal segmentation combining iterative self-training and a new data augmentation technique called Generative Blending Augmentation to realistically blend contrast-altered tumors to images in the target domain. The proposed method allowed to reduce the distribution shift between domains and realistically diversify tumor appearances, significantly improving segmentation performances. The proposed tumor alteration strategy can also be considered without a SinGAN tumor blending step, although results demonstrated consistent performance improvement with this refinement. Integrated within a conventional cross-modal segmentation pipeline (i.e. modality translation, then segmentation) for the MICCAI CrossMoDA 2022 challenge, the proposed approach allowed us to rank first on the vestibular schwannoma segmentation task. Potential improvements of our approach may include the introduction of confidence levels on pseudo-label accuracy to prevent the use of exceedingly noisy labels~\cite{liu2023cosst}, as well as evaluating more complex image alterations than tumor intensity rescaling. 

%
\bibliographystyle{IEEEtran}
\bibliography{IEEEabrv,bibliography}

\end{document}